\documentclass{elsarticle}

\usepackage{setspace}
\usepackage[margin=1in]{geometry}

\usepackage[utf8]{inputenc} 
\usepackage[T1]{fontenc}    
\usepackage{url}            
\usepackage{booktabs}       
\usepackage{amsfonts}       
\usepackage{nicefrac}       
\usepackage{microtype}      
\usepackage{lipsum}
\usepackage{graphicx}
\usepackage{caption}
\usepackage{subcaption}
\usepackage{algorithm}
\usepackage{algpseudocode}
\usepackage{amsmath}
\usepackage{amsfonts}
\usepackage{cleveref}

\journal{Computer-Aided Design}

\begin{document}

\begin{frontmatter}

\title{Learning topological operations on meshes with application to block decomposition of polygons}

\author[rvt1,rvt3]{A.~Narayanan\fnref{fn1}\corref{cor1}}
\ead{arjun.narayanan@berkeley.edu}

\author[rvt2,rvt3]{Y.~Pan\fnref{fn2}}
\ead{yllpan@berkeley.edu}

\author[rvt2,rvt3]{P.-O.~Persson\fnref{fn3}}
\ead{persson@berkeley.edu}

\address[rvt1]{Department of Mechanical Engineering, University of California, Berkeley, Berkeley, CA 94704, United States}
\address[rvt2]{Department of Mathematics, University of California, Berkeley, Berkeley, CA 94720, United States}
\address[rvt3]{Mathematics Group, Lawrence Berkeley National Laboratory, 1 Cyclotron Road, Berkeley, CA 94720, United States}
\cortext[cor1]{Corresponding author}
\fntext[fn1]{Graduate student, Department of Mechanical Engineering, University of California, Berkeley}
\fntext[fn2]{Graduate student, Department of Mathematics, University of California, Berkeley}
\fntext[fn3]{Professor, Department of Mathematics, University of California, Berkeley}


\begin{abstract}
We present a learning based framework for mesh quality improvement on unstructured triangular and quadrilateral meshes. Our model learns to improve mesh quality according to a prescribed objective function purely via self-play reinforcement learning with no prior heuristics. The actions performed on the mesh are standard local and global element operations. The goal is to minimize the deviation of the node degrees from their ideal values, which in the case of interior vertices leads to a minimization of irregular nodes. 
\end{abstract}

\end{frontmatter}

\section{Introduction} \label{sec:introduction}

Mesh generation is a crucial part of many applications, including the numerical simulation of partial differential equations as well as computer animation and visualization. While it can be discussed exactly what makes a mesh appropriate for a given situation, it is widely accepted that fewer number of irregular nodes lead to better quality meshes. Therefore, many mesh generation and mesh improvement methods have been proposed that aim to maximize the regularity of the mesh, in particular in the case of quadrilateral elements.

For triangular meshes, some of the most popular algorithms are the Delaunay refinement method \cite{shewchuk2002delaunay} and the advancing front method \cite{peraire1987adaptive}. The resulting meshes might be improved by local operations or smoothing, although typically based on element qualities rather than the regularity of the connectivities. Some quadrilateral mesh generators are also based on a direct approach, such as the paving method \cite{blacker1991paving}, but most are using an indirect approach of creating quadrilateral elements from a triangular mesh. These methods include the popular Q-Morph method \cite{owen1999qmorph}, element matching methods such as the Blossom-Quad method \cite{remacle2012blossom}, and so-called regularization or mesh simplification methods which improve an initial mesh using various mesh modification techniques \cite{daniels2008quadrilateral,tarini2010practical,akram2022structure,bommes2013quad}.

Although many of these mesh modification methods produce impressive results, we note that the algorithms for how they apply the various mesh operations are usually highly heuristic in nature \cite{docampo2020regularization,akram2022structure}. This is expected, since finding an optimal strategy is a complex discrete optimization problem. Therefore, in this work we explore the use of a deep neural network to learn optimal sequences of operations without human input. One of the main motivations behind this is that the problem fits well into the framework of reinforcement learning (RL) \cite{sutton2018reinforcement}, where the actions are the mesh operations and the rewards are the improvement of mesh regularity. The training can be performed in so-called self-play mode, where the policy is trained by learning to improve the connectivity of randomly generated meshes using a reward function that is proportional to the increase in mesh regularity.

In this work, we consider the case of planar straight-sided polygonal geometries. However, since our method is based purely on mesh connectivity, it may be applied to geometries with curved boundaries as well so long as the regularity of vertices on these boundaries is specified. We generate a coarse initial triangular mesh using the Delaunay refinement algorithm. In the case of quadrilateral meshes, we perform Catmull-Clark splits of the triangles, and we also introduce global mesh operations. One of these is the clean-up, which aims to reduce the total number of elements which is suitable for generation of block decompositions. 

A key component of our framework is the employment of the half-edge data structure, which in particular allows us to define a convolutional operation on unstructured meshes. A deep network is trained to produce a probability distribution for the various actions on local neighborhoods of the mesh, i.e., a policy. The policy is sampled to determine the next operation to perform. A powerful property of our method is that it generalizes to both triangular and quadrilateral meshes with minimal modifications to account for the different actions available on these meshes. We limit action selection to local mesh neighborhoods, allowing the learned policy to generalize well to a variety of mesh types and sizes that were not present in the training data. We demonstrate our methods on several polygonal shapes, where we consistently obtain meshes with optimal regularity. Extension of this method to arbitrary polygonal elements is reserved for future work.

Machine learning has been applied to numerous mesh generation problems before. Pointer networks \cite{vinyals2015pointer} have been used to generate convex hulls and Delaunay triangulations. Deep RL has been used to learn quadrilateral element extraction rules for mesh generation \cite{pan2021self,pan2023reinforcement}. RL has also been employed to learn adaptive mesh refinement strategies \cite{yang2023reinforcement,sluzalec2023quasi}. In \cite{diprete2023reinforcement}, RL was used to perform block decomposition of planar, straight-sided, axis aligned shapes using axis aligned cuts.

Our work differs from prior work in several key ways. Our objective function is purely based on the connectivity of the mesh and our framework aims to minimize the number of irregular vertices. We consider local topological edit operations as our action space. Our novel convolution operation on the half-edge data-structure provides a powerful, parameterized way of constructing state representations that encode neighborhood connectivity relationships. We employ a local neighborhood selection technique that allows us to generalize to different mesh sizes. These key features enable our method to work on both triangular and quadrilateral meshes of various sizes. 

The half-edge data-structure is able to represent arbitrary polygonal shapes in 2D. Thus our state representation method naturally extends to all such polygonal shapes. Our reinforcement learning framework can be applied to these shapes so long as an appropriate action space is defined. We hypothesize that the technique can be extended to 3-dimensions by leveraging the equivalent of the half-edge data-structure in higher dimensions \cite{dobkin1987primitives,dyedov2015ahf}. Prior work has explored the action space in 3D e.g. tetrahedral \cite{shewchuk2002two,klingner2007aggressive} and hexahedral meshes \cite{ledoux2010topological,tautgesa2003topology}.

\section{Problem Statement} \label{sec:problem-statement}

In the present work we are interested in optimizing the connectivity of triangular and quadrilateral meshes. The overall objective is to produce meshes where all the vertices have a specific number of incident edges. We refer to this as the \emph{desired degree} of a vertex. A vertex whose degree is the same as the desired degree is called \emph{regular}. A vertex whose degree is different from the desired degree is called \emph{irregular}, with the difference between the degree and the desired degree being a measure of the \emph{irregularity} of the vertex. Our framework allows the user to specify the desired degree on all vertices. The user is allowed to specify the desired degree of any newly introduced vertex. 

While there exist robust algorithms for triangular and quadrilateral meshing such as Delaunay triangulation and paving, these algorithms are not designed to produce meshes with a specific connectivity structure. A common approach is to use these algorithms as a starting point and improve the connectivity of the mesh through various topological mesh editing operations \cite{docampo2019towards}. We adopt this approach and frame our problem as a Markov Decision Process.

\subsection{Objective function} \label{sec:objective}

Consider a mesh with $N_v$ vertices. Let vertex $i$ have degree $d_i$ and desired degree $d^{*}_i$. Then its irregularity is $\Delta_i = d_i - d^{*}_i$. We compute a global score $s$ as the L1 norm of $s$, which is a measure of the total irregularity in the mesh.

\begin{align}
    s = \sum_{i=1}^{N_v} | \Delta_i \label{eq:global-score} |
\end{align}

\noindent Clearly, a mesh with all regular vertices will have a score $s = 0$.

\subsubsection{Heuristics to determine desired degree} \label{sec:tri-quad-heuristic}

Our heuristic for triangular (quadrilateral) meshes is based on achieving an interior angle of $60^{\circ}$ ($90^{\circ}$) in all elements. The desired degree of any vertex in the interior is 6 (4). The desired degree of a boundary vertex is chosen such that the average included angle in all elements incident on that boundary vertex is approximately $60^{\circ}$ ($90^{\circ}$) . The desired degree according to this heuristic can be expressed as,

\begin{align}
    d^{*} = \begin{cases}
        360 / \alpha &\text{interior vertex}\\
        \textrm{max} \left( \left \lfloor \theta / \alpha \right \rceil + 1, 2 \right) &\text{boundary vertex}
    \end{cases}  \label{eq:desired-degree}
\end{align}

\noindent where $\lfloor \cdot \rceil$ is the round to nearest integer operator, $\theta$ is the angle of the boundary at the vertex in question, and $\alpha$ is $60^{\circ}$ ($90^{\circ})$ for triangles (quadrilaterals). We observed that rounding to the nearest integer resulted in better performing models than using $d^{*}$ as a continuous value on the boundary. According to this heuristic, the desired degree of a new vertex introduced on the boundary is set to $4$ ($3$) since we assume that the edge on which the new vertex is introduced is a straight edge.

\subsection{Topological operations on meshes} \label{sec:topological}

\begin{figure}
    \centering
    \includegraphics[width=0.9\textwidth]{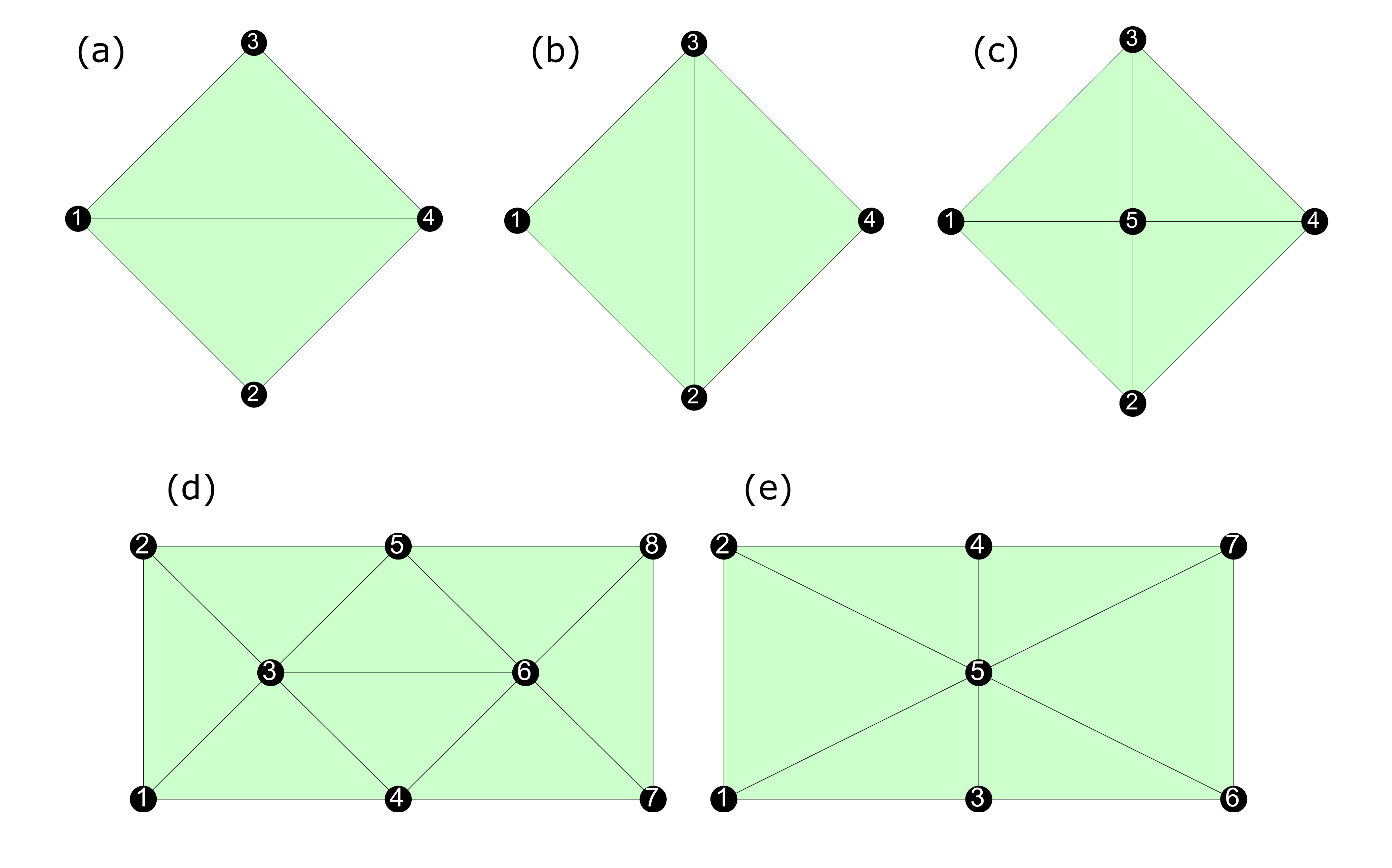}
    \caption{Configuration (a) and (b) are related by an edge flip. Configuration (c) can be produced by splitting the interior edge in either (a) or (b). Collapsing the edge between vertex 3-6 in (d) produces (e).}
    \label{fig:tri-ops}
\end{figure}

We define the following local operations on triangular meshes. See figure \cref{fig:tri-ops} for an illustration.

\begin{itemize}
    \item \textbf{Edge Flip:} An interior edge in a triangular mesh can be deleted and the resultant quadrilateral can be re-triangulated across its other diagonal. This can be seen as ``flipping'' an edge between two possible states.
    \item \textbf{Edge Split:} Any edge in a triangular mesh can be split by inserting a new vertex on the edge and connecting it to the opposite vertices in the adjacent triangles.
    \item \textbf{Edge Collapse:} An interior edge in a triangular mesh can be collapsed resulting in the deletion of the two triangles associated with this edge.
\end{itemize}

\begin{figure}
    \centering
    \includegraphics[width=0.9\textwidth]{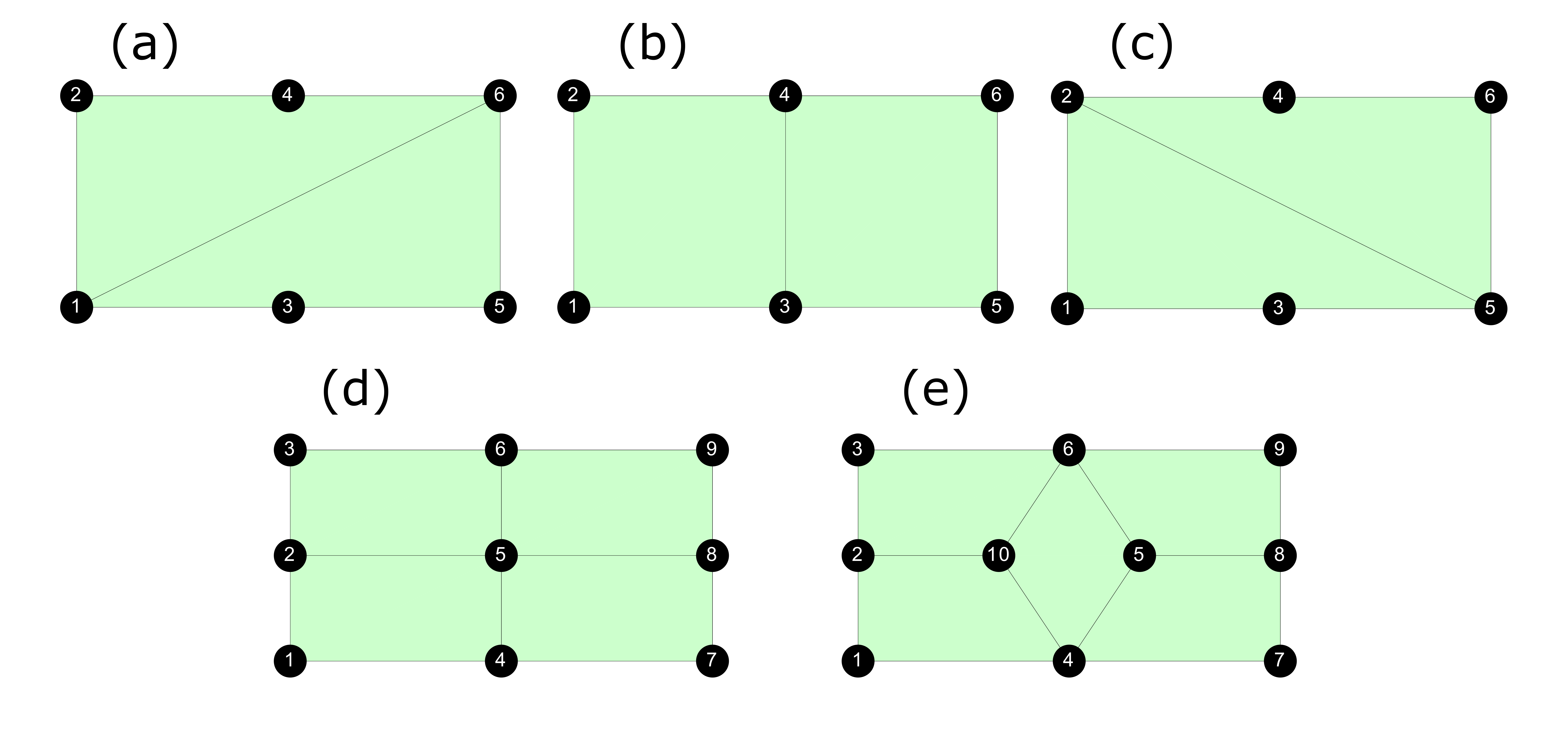}
    \caption{Configuration (a) and (c) can be obtained from (b) via an edge flip. Configuration (e) is obtained from (d) via a vertex split, and the operation can be reversed via an element collapse.}
    \label{fig:quad-ops}
\end{figure}

Similarly, we define the following local operations on quadrilateral meshes. See figure \cref{fig:quad-ops} for an illustration.

\begin{itemize}
    \item \textbf{Edge Flip:} An  interior edge in a quadrilateral mesh can be deleted, and the resultant hexagon can be quad-meshed in two new ways. This can be seen as ``flipping'' an edge clockwise or counter-clockwise.
    \item \textbf{Vertex Split:} A vertex in a quad mesh can be split along an interior edge incident at that vertex. This results in the insertion of a new vertex and a new element into the mesh.
    \item \textbf{Element Collapse:} A quadrilateral element can be collapsed along either diagonal by merging the two opposite vertices. The collapse operation can be seen as the inverse of the split operation defined above.
\end{itemize}

\begin{figure}
    \centering
    \includegraphics[width=0.9\textwidth]{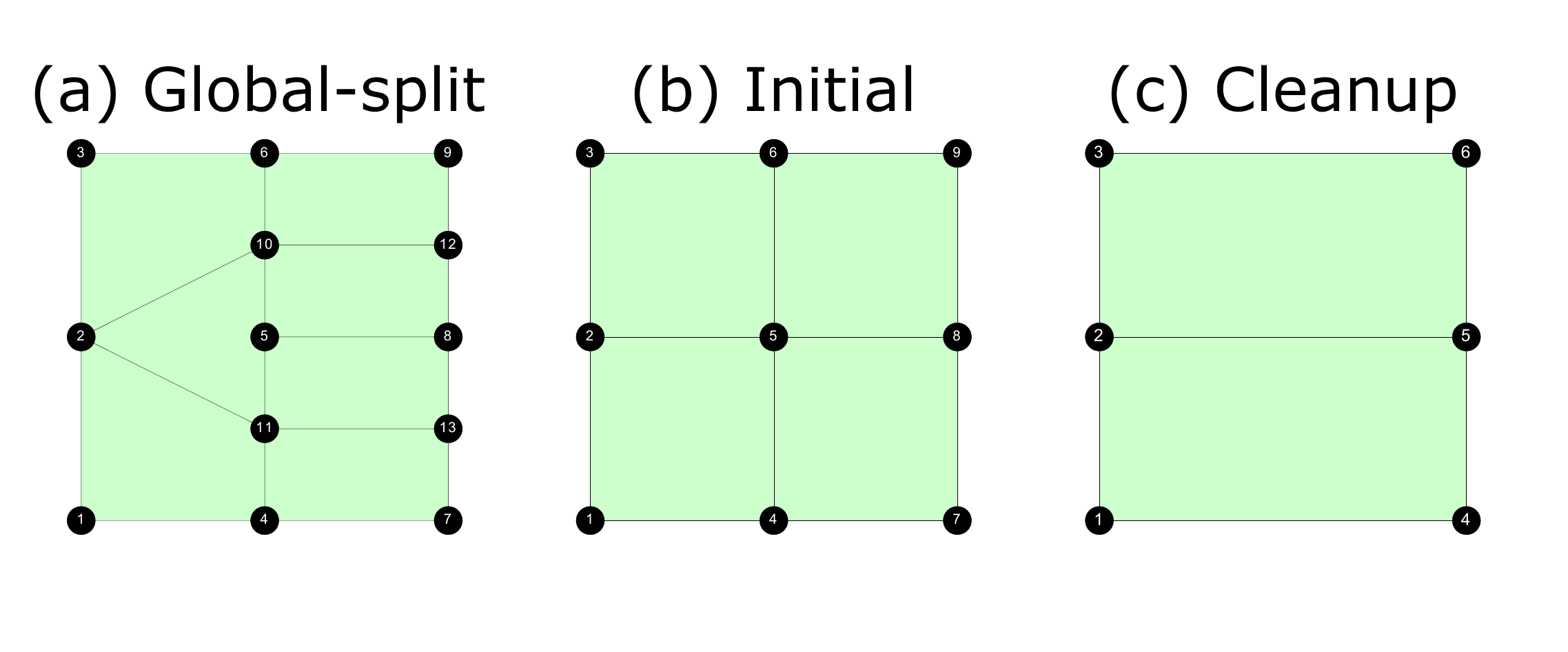}
    \caption{Performing a global split on the edge between vertices 2 and 5 in the initial mesh (b) produces the mesh in (a). Alternatively, the sequence of edges between vertices 4 -- 5 -- 6 in the initial mesh (b) can be deleted by merging the neighboring elements, resulting in configuration (c).}
    \label{fig:quad-global-ops}
\end{figure}

For quadrilateral meshes we also define the following global mesh editing operations. They are global in the sense that they can affect the topology of the mesh far away from where they are applied. See figure \cref{fig:quad-global-ops} for an illustration.

\begin{itemize}
    \item \textbf{Global Split:} This operation splits an edge by inserting a quadrilateral element and introducing vertices on the edges in the two adjacent quadrilateral elements. The introduced vertices are hanging vertices -- therefore we recover an all-quadrilateral mesh by propagating edges from the hanging vertices and sequentially splitting elements until the split terminates on a boundary.
    \item \textbf{Global Cleanup:} In some situations, global lines -- which represent a sequence of edges -- can be deleted by merging adjacent elements. The global line either terminates on the boundaries of the mesh or forms a closed loop. We currently handle the situation where the global line terminates on the boundaries. (For meshes representing closed surfaces it would be important to consider the case of closed loops.) This operation results in the deletion of a sequence of vertices and elements. Vertices are distinguished into geometric and non-geometric vertices. Geometric vertices are those vertices which are integral in defining the geometry -- these vertices cannot be deleted. The conditions under which we can perform this cleanup operation are (a) the end-points are on the boundary, are non-geometric, and have degree 3, (b) all interior vertices are non-geometric and have degree 4. The cleanup operation is a powerful operation since it simplifies the problem and brings irregular vertices closer together. This strategy is particularly relevant for block decomposition of polygonal shapes.
\end{itemize}

\section{Mesh Representation and Operations} \label{sec:mesh-representation}

\subsection{The half-edge data structure} \label{sec:halfedge}

We employ the doubly-connected edge list (DCEL), also known as the half-edge data-structure, to represent our meshes. The advantage of the DCEL is that (a) it enables efficient implementations of the mesh editing operations described in \cref{sec:topological}, and (b) we utilize fundamental DCEL operations to represent the local topology in a given mesh region which is important to determine the appropriate action to be applied. The DCEL can be used to represent any planar, manifold mesh and as such allows our method to work on all such meshes. Extensions to the DCEL have been developed for non-manifold meshes and 3D volumetric meshes \cite{dobkin1987primitives, dyedov2015ahf}.

\begin{figure}
    \centering
    \includegraphics[width=0.3\textwidth]{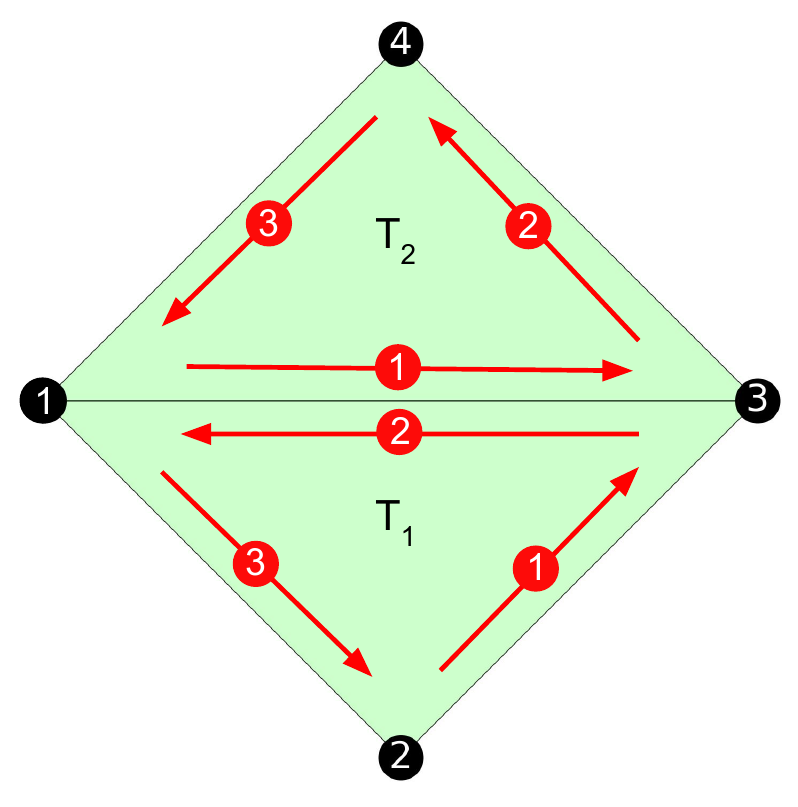}
    \caption{Representing two triangular elements using the DCEL. The half-edges in each element are shown as red arrows. Each half-edge contains a pointer to the counter-clockwise \emph{next} half-edge in the same element e.g. in triangle $T_2$, half-edge 2's  \emph{next} pointer points to half-edge 3. Half-edges in the interior of the mesh have a \emph{twin} pointer to the half-edge in the adjacent element e.g. the \emph{twin} of half-edge 1 in triangle $T_2$ is half-edge 2 in triangle $T_1$. We additionally associate each half-edge with a unique vertex in the element. For triangles we associate the vertex opposite a given half-edge e.g. half-edge 1 in triangle $T_2$ is associated with vertex $4$. For quadrilaterals we associate the vertex at the origin of the half-edge.}
    \label{fig:half-edge-ds}
\end{figure}

Briefly, the DCEL exploits the fact that each mesh edge is shared by exactly two mesh elements (except on the boundary). The DCEL represents each mesh edge as a pair of oriented half-edges pointing in opposite directions. Each half-edge contains a pointer to the counter-clockwise \emph{next} half edge in the same element, and a pointer to the \emph{twin} half-edge in the adjacent element. Each element contains a pointer to one of its half-edges (chosen arbitrarily) which induces an ordering on the half-edges in an element. Elements can be ordered by their global index in the mesh -- this induces a global ordering on half-edges in the mesh. Each half-edge may be associated with a unique vertex, e.g. the vertex at the origin of the half-edge. See fig. \ref{fig:half-edge-ds} for an illustration. Further details about the DCEL can be found in a standard resource on computational geometry, for example \cite{mark2008computational}. 

\subsection{Algorithmic complexity and parametrization of mesh editing operations}

All of the local editing operations defined in \cref{sec:topological} for triangles and quadrilaterals can be executed using the DCEL in constant time (assuming an upper bound on the maximum degree of a vertex). This is a powerful advantage offered by the DCEL compared to other mesh representations. For instance, when flipping a particular half-edge it is important to know which are the two neighboring elements across that edge -- this is readily available in the DCEL. 

The global operations defined for quadrilateral meshes in \cref{sec:topological} requires connectivity editing operations that can propagate through several elements of the mesh before terminating. The algorithm for these operations scales linearly with the size of the mesh. In particular we disallow situations where global splits may result in the formation of loops or that do not terminate in a fixed number of iterations proportional to the size of the mesh. For the cleanup, we observe that every half-edge either lies on a cleanup path or does not. Performing a cleanup on a path does not affect the ability to cleanup other paths. Therefore all cleanups possible in a mesh can be performed by visiting every half-edge exactly once.

Our framework optimizes a policy to perform sequences of mesh editing operations to achieve a given objective. All operations other than the global-cleanup are valid operations that can be learned by the policy. Whenever a global-cleanup is valid, it is always performed. We choose to do this because the cleanup simplifies the problem size and brings irregular vertices together making it easier to improve the connectivity of the mesh. A cleanup only deletes regular vertices according to our heuristic and never introduces any new irregular vertices in the mesh. Further, the cleanup is very useful in performing block decompositions of polygons.

We parametrize all the mesh editing operations in terms of half-edges. In a given mesh, specifying a particular half-edge and a particular type of edit determines an operation on the mesh. We have 3 operations per half-edge in the case of triangular meshes -- flip, split, and collapse. Further, we have 5 operations per half-edge in the case of quadrilateral meshes -- right-flip, left-flip, split, collapse, and global-split. There is some redundancy in this representation of actions on the mesh. For instance, flipping a half-edge and its twin are equivalent operations. We choose to retain this redundancy because (a) it fits in well with our half-edge framework, (b) the size of the state representation is larger only by a constant factor, and (c) it exposes the symmetries in the half-edge representation and may be seen as data augmentation in our state representation leading to more robust learning. Further, some actions -- like the quadrilateral split -- are not equivalent when performed on a half-edge and its twin.

\section{Formulation as a Reinforcement Learning Problem}

\subsection{Constructing the reward function} \label{sec:solution}

Clearly, a mesh with all regular vertices will have a score $s = 0$. Under the assumption of the heuristic described in \cref{sec:tri-quad-heuristic}, all the topological edit operations described in \cref{sec:topological} are \emph{zero-sum} leaving the quantity $s^{*} = | \sum_{i} \Delta_i |$ invariant for a given mesh. This does not hold true if we change the heuristic for the desired degree of newly introduced vertices from what we described in \cref{sec:tri-quad-heuristic}. If a mesh contains irregular vertices all of the same sign then its global score \cref{eq:global-score} cannot be improved. $s^{*}$ provides a lower bound on the score $s$,

\begin{align}
    s^{*} = \left| \sum_{i}^{N_v} \Delta_i \right| \leq \sum_{i}^{N_v} | \Delta_i | = s \label{eq:objective-lower-bound}
\end{align}

We call $s^{*}$ the \emph{optimum score}. It is not clear if a score $s^{*}$ can always be attained for a given mesh, however it serves as a useful measure of performance. The goal of our reinforcement learning framework is to learn sequences of actions that minimize $s$ for a given mesh. In particular, consider a mesh $M_t$ with score $s_t$ at some time $t$. We now perform a mesh editing operation $a_t$ on it to obtain mesh $M_{t+1}$ with score $s_{t+1}$. Our agent is trained with reward $r_t$,

\begin{align}
    r_t = s_t - s_{t+1} \label{eq:reward}
\end{align}

An agent starting with an initial mesh $M_1$ transformed through a sequence of $n$ operations $a_1, a_2, \ldots a_n$ collects reward $r_1, r_2, \ldots r_n$. We consider the discounted return from state $M_t$ as,

\begin{align}
    G_{t} = \sum_{k=t}^{n} \gamma^{k-t} r_k \label{eq:return}
\end{align}

\noindent with discount factor $\gamma$. Observe that the maximum possible return from this state is $G^{*} = s_t - s^*$. Thus, we consider the normalized return $\overline{G_t}$ as the \emph{advantage} function to train our reinforcement learning agent,

\begin{align}
    \overline{G_t} = \frac{G_t}{s_t - s^{*}} \label{eq:advantage}
\end{align}

\noindent The return \cref{eq:return} collected on meshes of different sizes will be different simply because larger meshes tend to have more irregularities. By normalizing the return in \cref{eq:advantage}, we ensure that actions are appropriately weighted during policy optimization. The mesh environment terminates when the mesh score $s_t = s^*$ or when a given number of mesh editing steps have been taken. We choose the maximum number of steps to be proportional to the number of mesh elements in the initial mesh.

While our current experiments are based on the objective described above, one could consider other measures -- such as element quality -- as an objective. This would require a consideration of geometry and topology and is reserved for future work.

\subsection{Convolution operation on the DCEL data-structure} \label{sec:dcel-convolution}

All of the actions, apart from the global-cleanup, affect the topology of the mesh locally. In order to determine if an action produces desirable outcomes in a particular neighborhood of the mesh, we need to understand the topology of this neighborhood. We require a representation of the local topology around each half-edge in order to select a suitable operation. In the language of reinforcement learning, this representation of the local topology is the \emph{state} of a half-edge. The connectivity information in the immediate neighborhood of a half-edge is most relevant to determine the appropriate action to take in this neighborhood. We present here a convolution operation on the DCEL data-structure that encodes topological information around every half-edge. Indeed, this operation may be interpreted as a convolution on the graph induced by the half-edge connectivity. Iterative application of this convolution encodes topological information in a growing field-of-view around every half-edge. Further, this convolution operations can be efficiently implemented on modern GPU hardware.

Determining the appropriate action to take on a given half edge requires us to inspect the degree and irregularity of vertices in a neighborhood around the half-edge. Since the meshes we consider are unstructured, it is not immediately obvious which vertices to consider and in what order to consider them in. Our key observation is that the fundamental DCEL operations can be leveraged to construct a state representation for each half-edge that has a specific ordering. Our convolution operation requires two fundamental pieces of information both of which are easily available from the DCEL. For each half-edge we need to know the indices of (a) all the cyclic-next half-edges from the given element, and (b) the twin half-edges from the neighboring element. (a) is easily achieved by using the \texttt{next} operation repeatedly -- 3 for triangles and 4 for quadrilaterals. (b) is fundamentally part of the DCEL data-structure.

As described in \cref{sec:halfedge}, there is a natural global ordering for all the half-edges in the mesh. Half-edges from the same element appear sequentially in this global ordering. If the half-edges are stored in this order, the cycle operation can be implemented efficiently as a sequence of matrix \texttt{reshape} operations which are provided by most array based programming languages. Consider a mesh with $N_h$ half-edges with the state of each half-edge represented by an $N_f$ dimensional vector. This data when stored in sequential order can be represented by a matrix $x \in \mathbb{R}^{N_f \times N_h}$. \Cref{alg:tri-cycle} describes the cycle operation applied to this state matrix for triangular meshes. The extension to quadrilateral meshes or other polygonal meshes is straightforward. (We assume that n-dimensional arrays are stored in column-major order. We adopt a syntax that closely follows the Julia/MATLAB Programming Language.)

\begin{algorithm}
\caption{Cycle operation on triangular meshes} \label{alg:tri-cycle}
    \hspace*{\algorithmicindent} \textbf{Input} \texttt{x} $\in \mathbb{R}^{N_f \times N_h} $ \\
    \hspace*{\algorithmicindent} \textbf{Output} \texttt{y} $\in \mathbb{R}^{3N_f \times N_h}$
\begin{algorithmic}
\State  \texttt{x} $\gets$ \texttt{reshape(x, N\textsubscript{f}, 3, :)}
\State \texttt{x1} $\gets$ \texttt{reshape(x, 3N\textsubscript{f}, 1, :)}
\State \texttt{x2} $\gets$ \texttt{reshape(x[:, [2, 3, 1], :], 3N\textsubscript{f}, 1, :)}
\State \texttt{x3} $\gets$ \texttt{reshape(x[:, [3, 1, 2], :], 3N\textsubscript{f}, 1, :)}
\State \texttt{y} $\gets$ \texttt{concatenate x1, x2, and x3 along the second dimension (i.e. columns)}
\State \texttt{y} $\gets$ \texttt{reshape(y, 3N\textsubscript{f}, :)}
\end{algorithmic}
\end{algorithm}

Information from twin half-edges is easily obtained by selecting the appropriate columns from the feature matrix. We use a learnable vector as the twin feature for edges on the boundary. Our basic convolution operation involves cycling the current feature matrix, obtaining the features from the twin half-edges, and concatenating all of the features together. The resultant matrix is processed by a linear layer, followed by normalization and a non-linear activation function. We include multiple such blocks in our model. Under the operation of each block, every half-edge receives information from all the half-edges within the same element and the twin half-edge from the adjacent element. After repeated application of such blocks, the final feature matrix will contain an encoding of the local topology in a field-of-view around every half-edge. The size of this field of view grows linearly with the number of blocks.

The initial feature matrix fed to this block is $x_0 \in \mathbb{R}^{2 \times N_h}$. Recall from \cref{sec:halfedge} that each half-edge is associated with a vertex. The initial feature matrix consists of the degree and irregularity of the associated vertices for every half-edge. This initial feature matrix is projected to a high dimensional space on which the convolution described above is applied. The final layer projects the features into an $N_a \times N_h$ matrix where $N_a$ is the number of actions per half-edge.

\begin{figure}
    \centering
    \includegraphics[width=\textwidth]{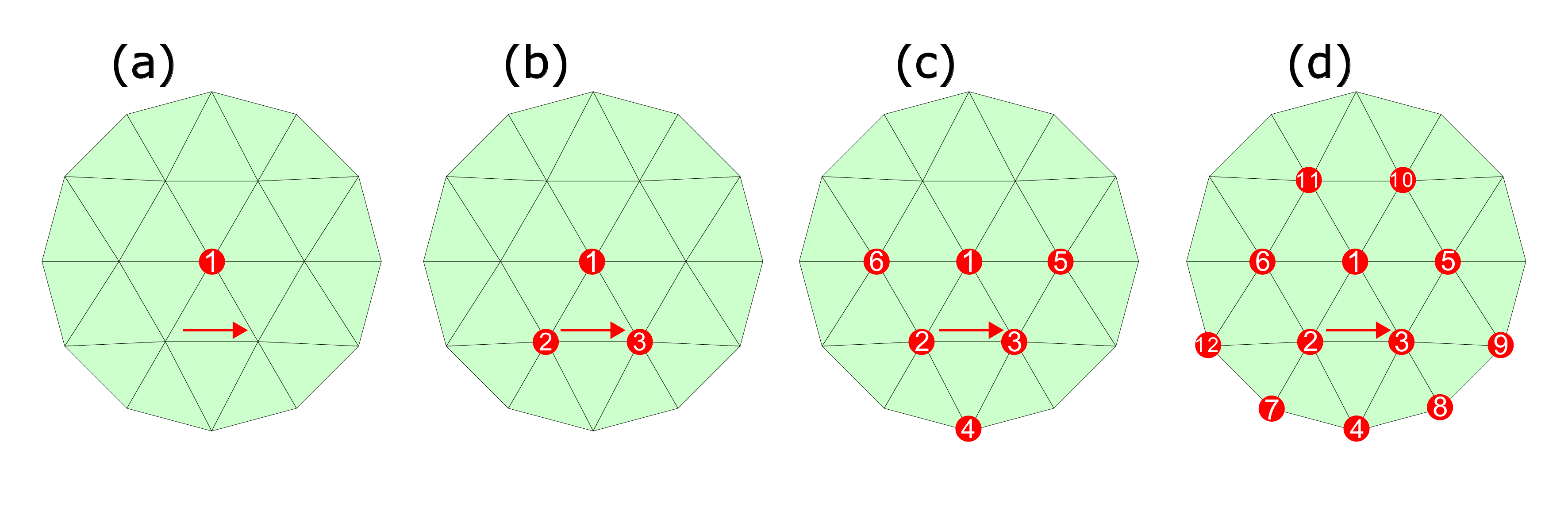}
    \caption{Repeated application of the convolution operation produces an increasing field of view around every half-edge. For triangles, we associate every half-edge with the vertex opposite it in the same triangle (fig. a). A \texttt{cycle} operation gathers information from the remaining vertices in the element (fig. b). Repeated application of \texttt{twin} and \texttt{cycle} produces the ordered list of vertices in fig. (c) and (d).}
    \label{fig:tri-template}
\end{figure}

\subsubsection{Action selection by the agent}

The size of meshes can vary as the agent manipulates the mesh. The total number of actions available to the agent varies with the size of the mesh. To ensure that the agent can generalize across different mesh sizes, we found it important that our policy is represented by a fixed sized vector representing the probabilities of selecting various actions. 

To do this, we generate a list of vertices which we call the \emph{template} around each half-edge (see \cref{fig:tri-template}.) The template can be constructed using operations similar to the convolution described in \cref{sec:dcel-convolution}. We use dummy vertices if the template goes outside the boundary of the mesh. We then compute the score \cref{eq:global-score} restricted to each template. This is a measure of the local irregularity around every half-edge. Action selection is then restricted to the half-edges in the template with the highest local score with ties broken randomly. Thus we consider an $N_a \times N_l$ subset of the output feature matrix from \cref{sec:dcel-convolution} where $N_l$ is the number of half-edges in the template. This matrix is flattened and passed through a softmax layer to obtain a probability distribution over actions in the template. We sample from this distribution to take a step into a new mesh state.

\subsubsection{Training the agent in self-play}

The initial states for self-play are randomly generated polygonal shapes. We randomize the degree of polygon between set bounds. We perform Delaunay refinement meshing of this shape and use that as the input to the triangular mesh agent. For the quadrilateral agent, we perform the Catmull-Clark splits on the triangles to get an all quad initial mesh.

The agent is allowed to interact with the mesh and perform operations on it for a finite number of steps or until the agent achieves the optimum score $s^{*}$ whichever comes first. We generate multiple rollouts of the current policy, and train the agent using the Proximal Policy Optimization (PPO) algorithm \cite{schulman2017proximal}. This data collection and policy improvement step is repeatedly performed to learn an optimal policy. We use \cref{eq:advantage} as the advantage function in the PPO algorithm. We add an entropy regularization to the loss function to avoid local minima and balance exploration with exploitation.

\section{Results}

\subsection{Triangular Meshes} \label{sec:tri-results}

\begin{figure}
     \centering
     \begin{subfigure}[b]{0.48\textwidth}
         \centering
         \includegraphics[width=\textwidth]{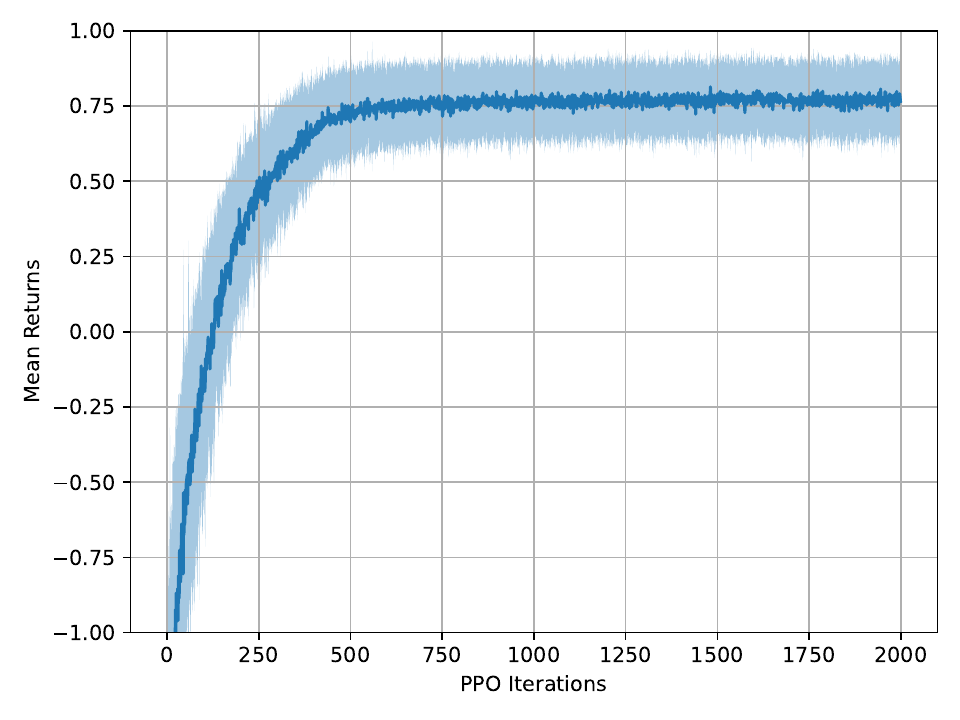}
         \caption{Average performance over training history} \label{fig:tri-ppo-returns}
     \end{subfigure}
     \hfill
     \begin{subfigure}[b]{0.48\textwidth}
         \centering
         \includegraphics[width=\textwidth]{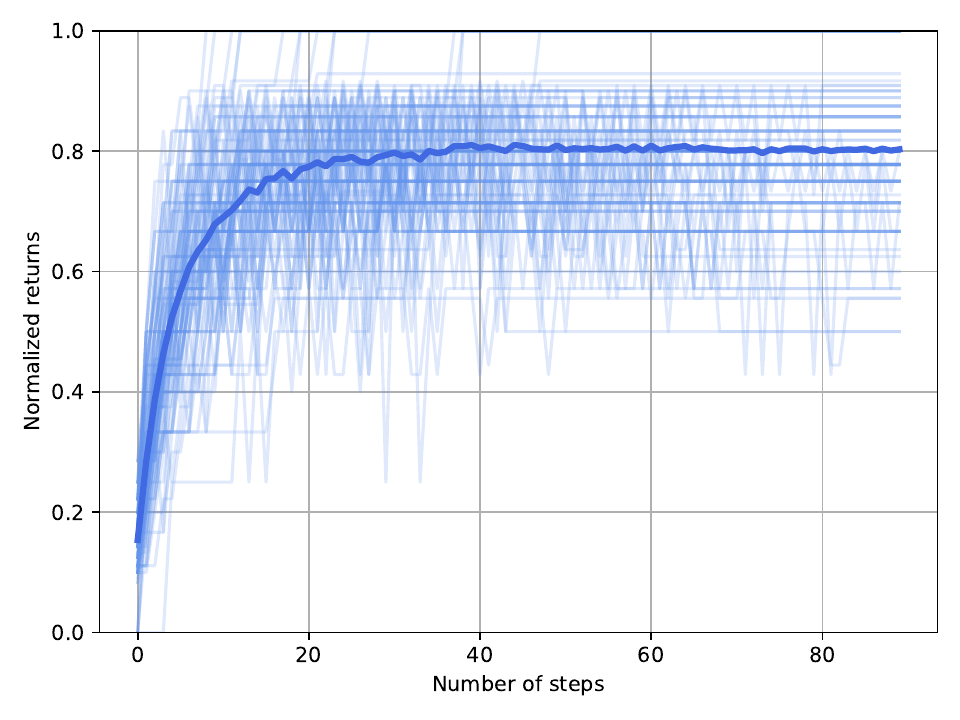}
         \caption{Evaluating the trained agent on multiple rollouts} \label{fig:tri-return-vs-steps}
     \end{subfigure}
     \hfill
        \caption{(a) Performance of the triangle mesh agent over the training history. Solid line represents the average normalized return over 100 meshes evaluated periodically during training. Shaded region represents the 1-standard deviation envelope. (b) Performance of the trained agent over 100 rollouts. The agent incrementally improves mesh quality up to a certain number of steps. Notice that returns do not increase monotonically, indicating that a greedy strategy may not be effective for this problem.} \label{fig:tri-returns}
\end{figure}

The triangular mesh agent was trained on random shapes consisting of 10 to 30 sided polygons. The initial mesh was a Delaunay refinement mesh generated by the Triangle package \cite{shewchuk2002delaunay}. \Cref{fig:tri-returns} shows the learning curves of our agent over training history, and the performance of the trained model over 100 rollouts. The average normalized single-shot performance over 100 meshes was about 0.81 ($\sigma = 0.11$). However, since the learned policy is stochastic, a simple way to improve the performance is to run the policy $k$ times from the same initial state and pick the best mesh. Using $k = 10$ samples per mesh and averaging over 100 random meshes, the performance improved to 0.86 ($\sigma = 0.08$).

\Cref{tab:tri-generalize} demonstrates the generalization capability of the learned policy. By using a fixed sized local template, the same agent can be evaluated on meshes of various sizes with good results. We do observe some reduction in model performance on larger meshes. Irregularities tend to be separated by greater distances on larger meshes, requiring longer sequences of operations to effectively remove them. We believe that this is a major cause of performance reduction.

\begin{table}
\centering
\begin{tabular}{c|c|c}
     Polygon degree & Average & Standard deviation  \\ \hline
        3 - 10 & 0.83 & 0.19 \\
        10 - 20 & 0.87 & 0.08 \\
        20 - 30 & 0.83 & 0.10 \\
        30 - 40 & 0.78 & 0.08 \\
        40 - 50 & 0.75 & 0.07
\end{tabular}
\caption{Evaluating the triangle mesh agent on various sized random polygons. The agent was trained purely on 10 - 30 sided polygons but is able to generalize to other polygon sizes with minor deterioration in performance. The agent was evaluated by picking the best of 10 rollouts per geometry, with the statistics computed over 100 randomly generated shapes. The results demonstrate the effectiveness of using a fixed-sized local template which enables better generalization to different mesh sizes.} \label{tab:tri-generalize}
\end{table}

\begin{figure}
    \centering
    \begin{subfigure}[b]{.3\linewidth}
    \includegraphics[width=\linewidth]{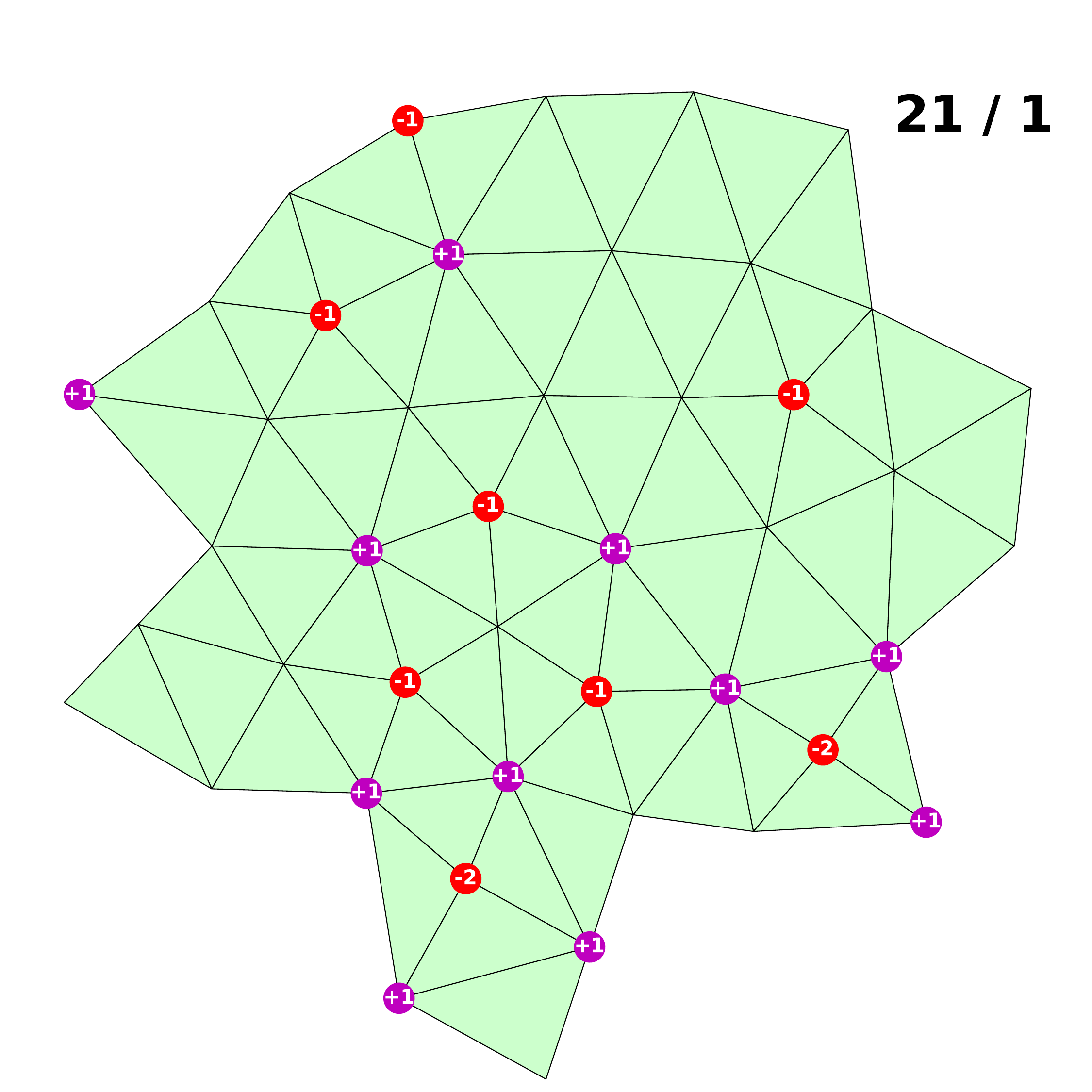}
    \caption{}
    \end{subfigure}
    \begin{subfigure}[b]{.3\linewidth}
    \includegraphics[width=\linewidth]{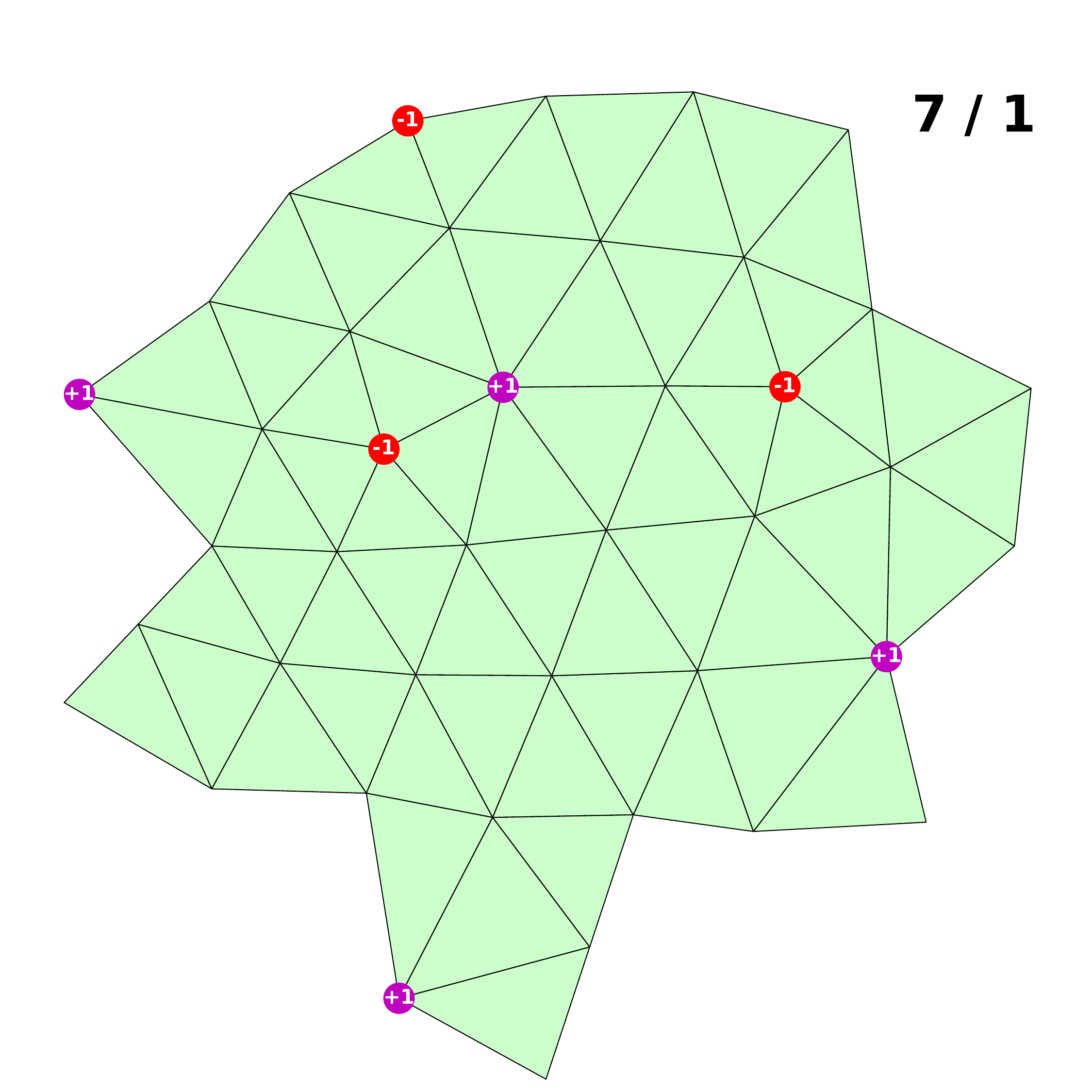}
    \caption{}
    \end{subfigure}
    \begin{subfigure}[b]{.3\linewidth}
    \includegraphics[width=\linewidth]{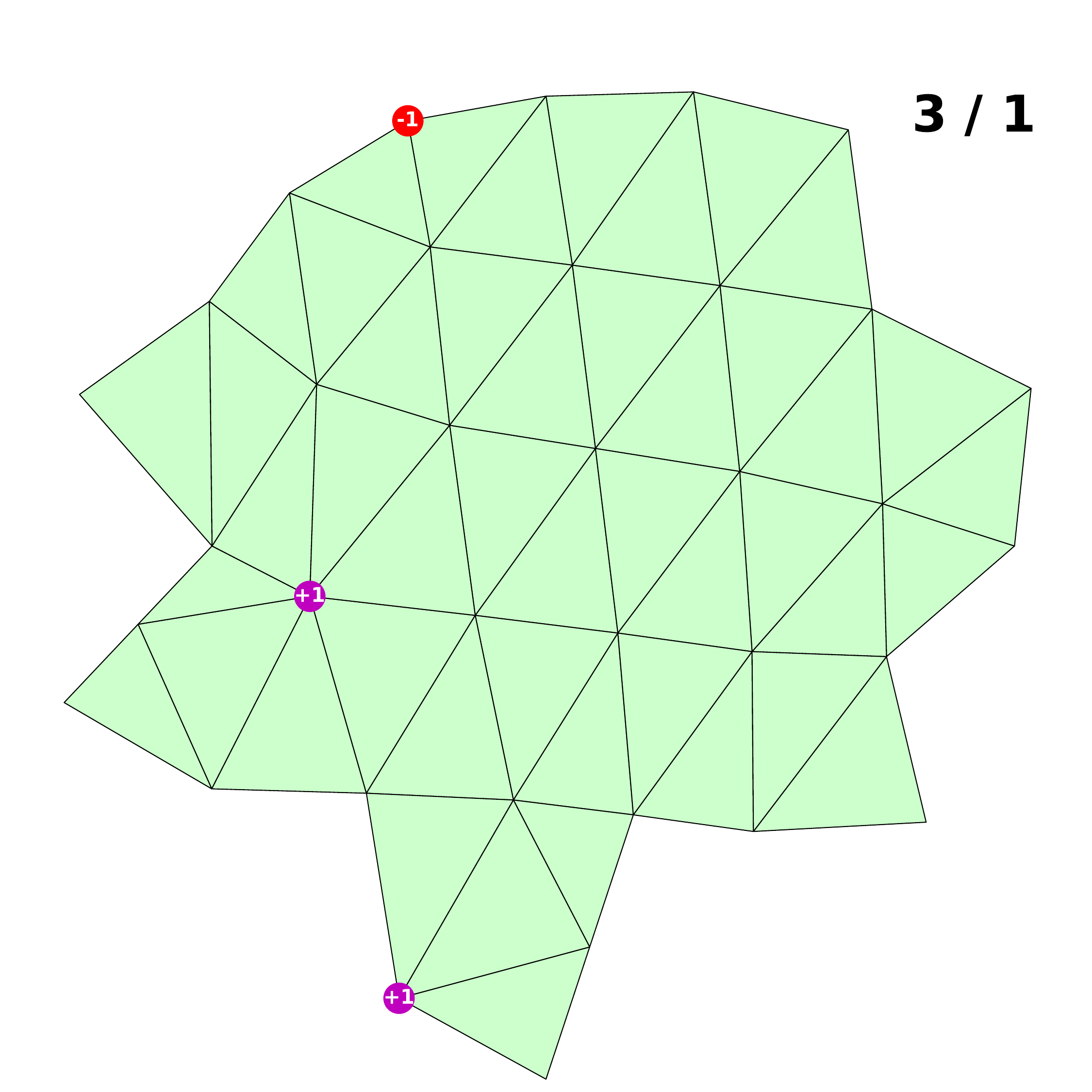}
    \caption{}
    \end{subfigure}
    \caption{Example rollout of the triangular mesh agent on a 20-sided polygon. Irregular vertices are marked in color, with the current score and optimum score shown at the top right for each figure. (a) is the initial Delaunay refinement mesh (b) is at an intermediate stage and (c) is the final mesh after 27 operations.} \label{fig:tri-20-side}
\end{figure}

\subsection{Quadrilateral Meshes} \label{sec:quad-results}

\begin{figure}
     \centering
     \begin{subfigure}[b]{0.48\textwidth}
         \centering
         \includegraphics[width=\textwidth]{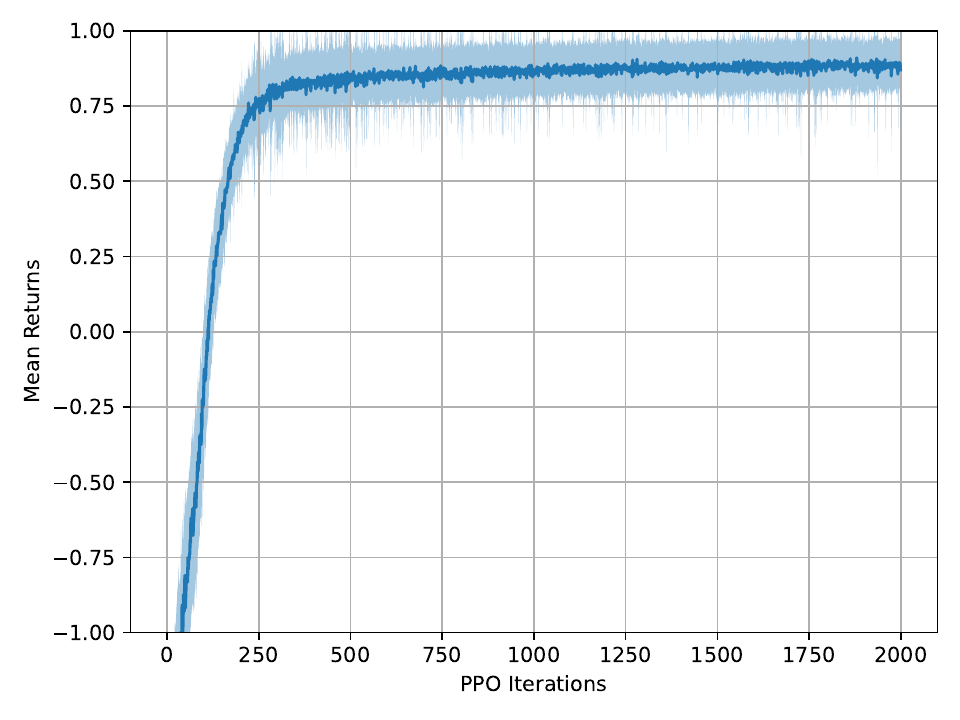}
         \caption{Average performance over training history}
         \label{fig:quad-ppo-returns}
     \end{subfigure}
     \hfill
     \begin{subfigure}[b]{0.48\textwidth}
         \centering
         \includegraphics[width=\textwidth]{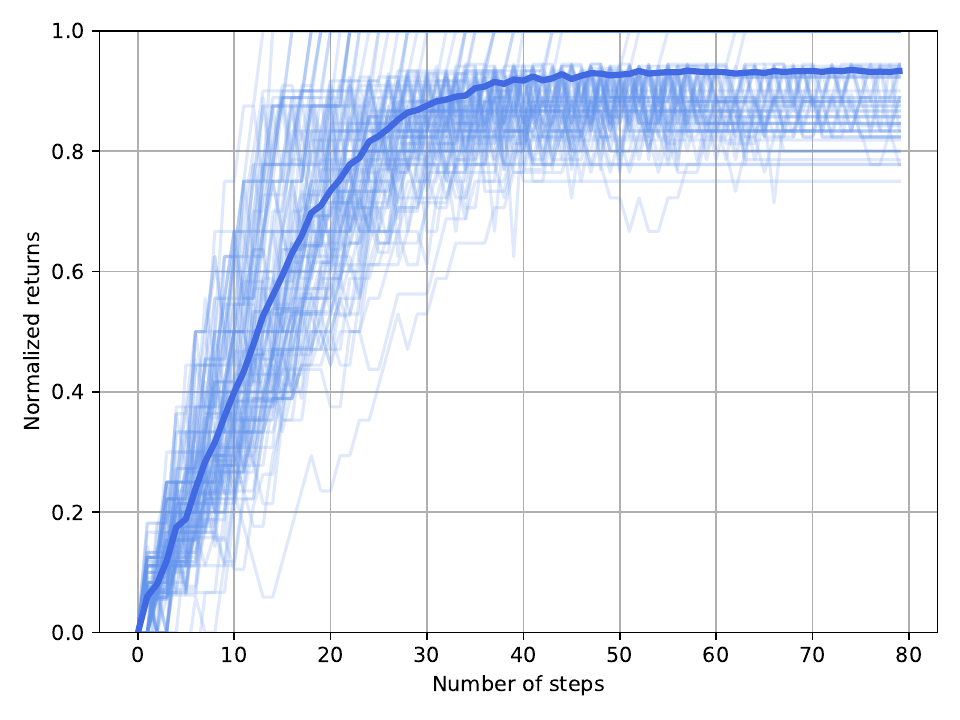}
         \caption{Evaluating the trained agent on multiple rollouts}
         \label{fig:quad-return-vs-steps}
     \end{subfigure}
     \hfill
        \caption{(a) Performance of the quadrilateral mesh agent over the training history. Solid line represents the average normalized returns evaluated over 100 meshes. Shaded region represents the 1-standard deviation envelope. The curve demonstrates that the agent is able to achieve good performance quite rapidly, and the learning remains stable over many training iterations. (b) Evaluating the trained model over multiple rollouts. Solid line represents the average performance of 100 rollouts. The graph demonstrates that increasing the maximum number of operations available to the agent has a big impact on performance initially, but only up to a certain point. Returns do not increase monotonically, highlighting that a greedy strategy may not be effective in this setting.}
        \label{fig:quad-returns}
\end{figure}

The quadrilateral mesh agent was trained on random shapes consisting of 10 to 30 sided polygons. \cref{fig:quad-ppo-returns} shows the average normalized returns over training for the quadrilateral mesh agent. We observe that the agent quickly learns operations that significantly improves the connectivity of the mesh to nearly optimal. Performance was assessed periodically during training by evaluating the model on 100 randomly generated meshes. \Cref{fig:quad-return-vs-steps} shows the evaluation of the best performing model on 100 trajectories. We observe that performance depends on the maximum number of steps given to the agent up to a certain point. The average normalized single-shot performance over 100 meshes was about 0.95 ($\sigma = 0.05$.). Using $k = 10$ samples per mesh and averaging over 100 random meshes, the performance improved to 0.992 ($\sigma = 0.02$).

Since our state representation is a fixed-sized local template around a half-edge of interest, our model generalizes well to polygons that were not part of the training dataset. \Cref{tab:quad-generalize} shows the performance of a model trained on 10 - 20 sided polygons that is able to generalize to larger sized polygonal shapes. We do observe some drop in the performance of the agent when mesh sizes are increased. In larger meshes, there is a greater chance of irregularities being separated from each other requiring longer range sequences of operations to remove these irregularities. We believe that this is a major cause of performance reduction.

\begin{figure}
    \centering
    \begin{subfigure}[b]{.3\linewidth}
    \includegraphics[width=\linewidth]{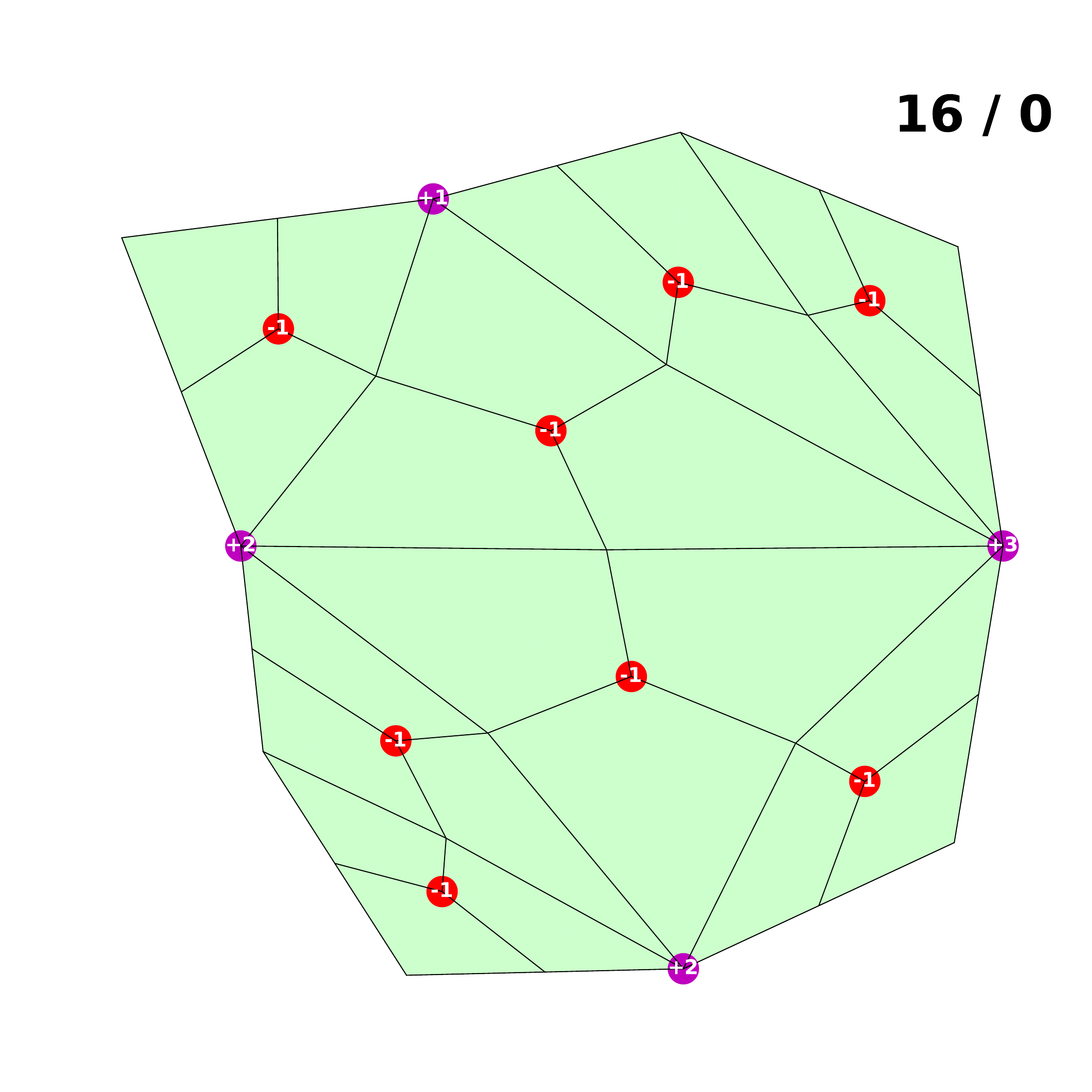}
    \caption{}
    \end{subfigure}
    \begin{subfigure}[b]{.3\linewidth}
    \includegraphics[width=\linewidth]{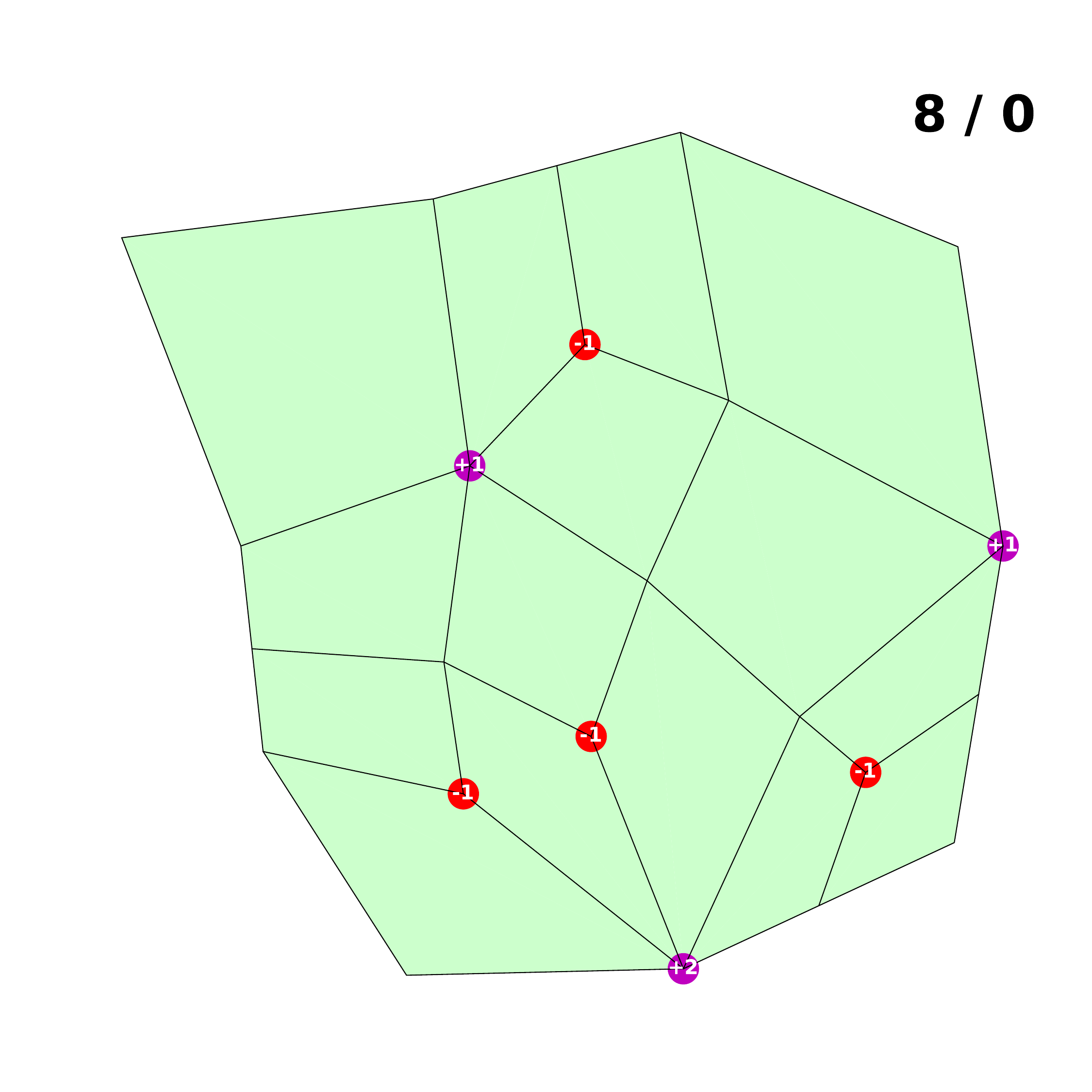}
    \caption{}
    \end{subfigure}
    \begin{subfigure}[b]{.3\linewidth}
    \includegraphics[width=\linewidth]{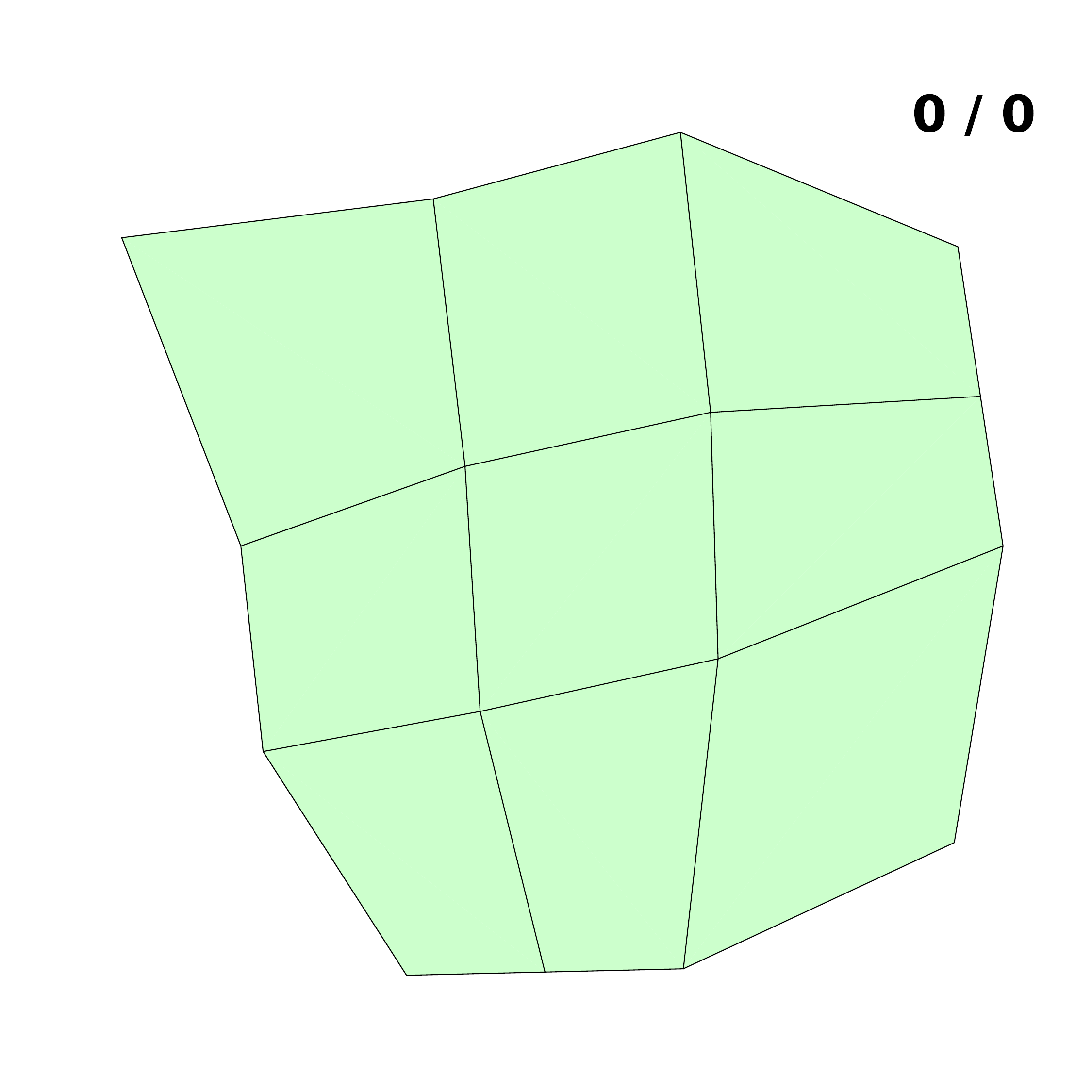}
    \caption{}
    \end{subfigure}
    \caption{Example rollout for a 10-sided polygon. Irregular vertices are marked in color. The mesh score and optimal score are shown at the top right for each figure. (a) is the initial mesh after Delaunay triangulation and catmull-clark splits, (b) is at an intermediate stage, and (c) is the final mesh after 18 operations.} \label{fig:quad-10-side}
\end{figure}

\begin{figure}
    \centering
    \begin{subfigure}[b]{.3\linewidth}
    \includegraphics[width=\linewidth]{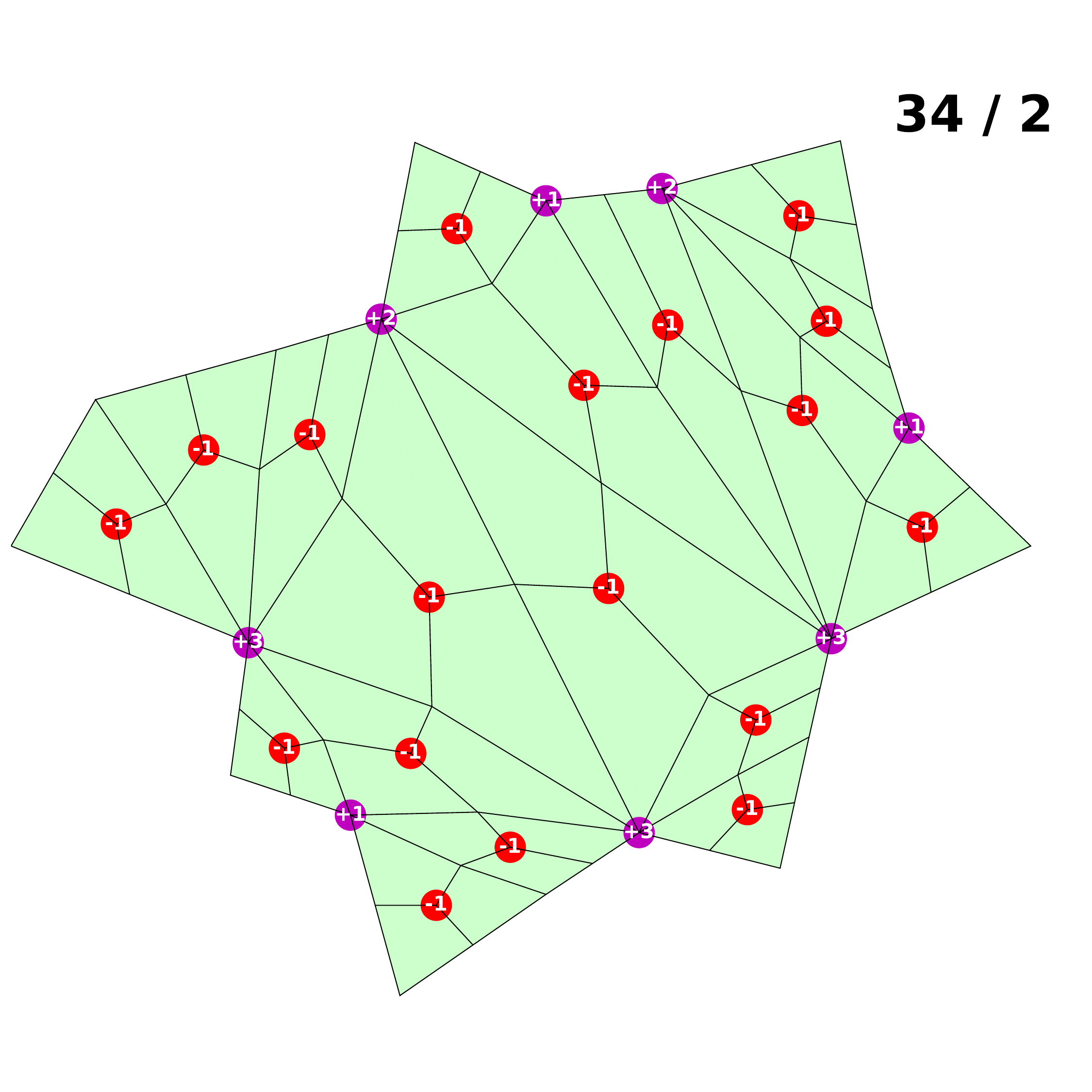}
    \caption{}
    \end{subfigure}
    \begin{subfigure}[b]{.3\linewidth}
    \includegraphics[width=\linewidth]{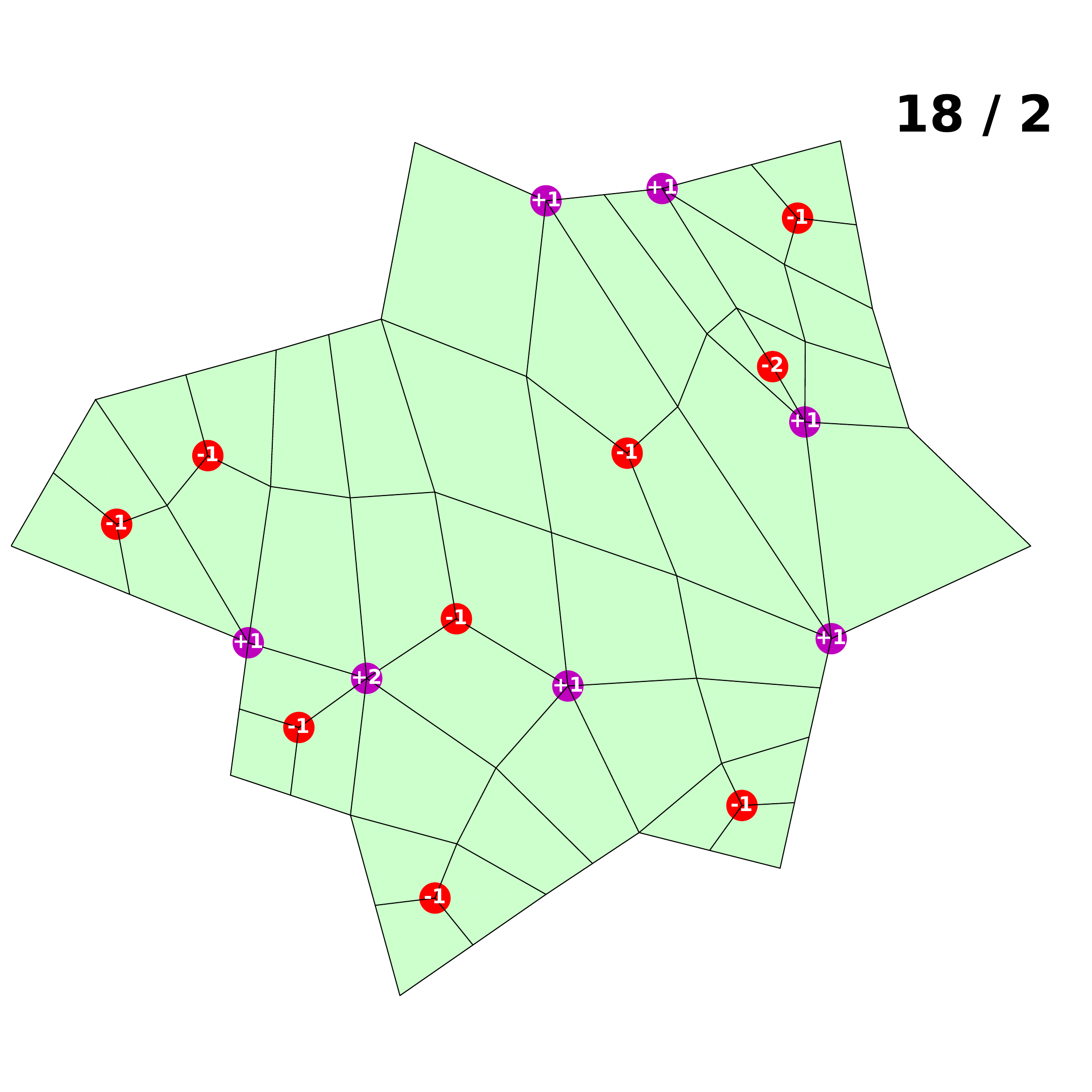}
    \caption{}
    \end{subfigure}
    \begin{subfigure}[b]{.3\linewidth}
    \includegraphics[width=\linewidth]{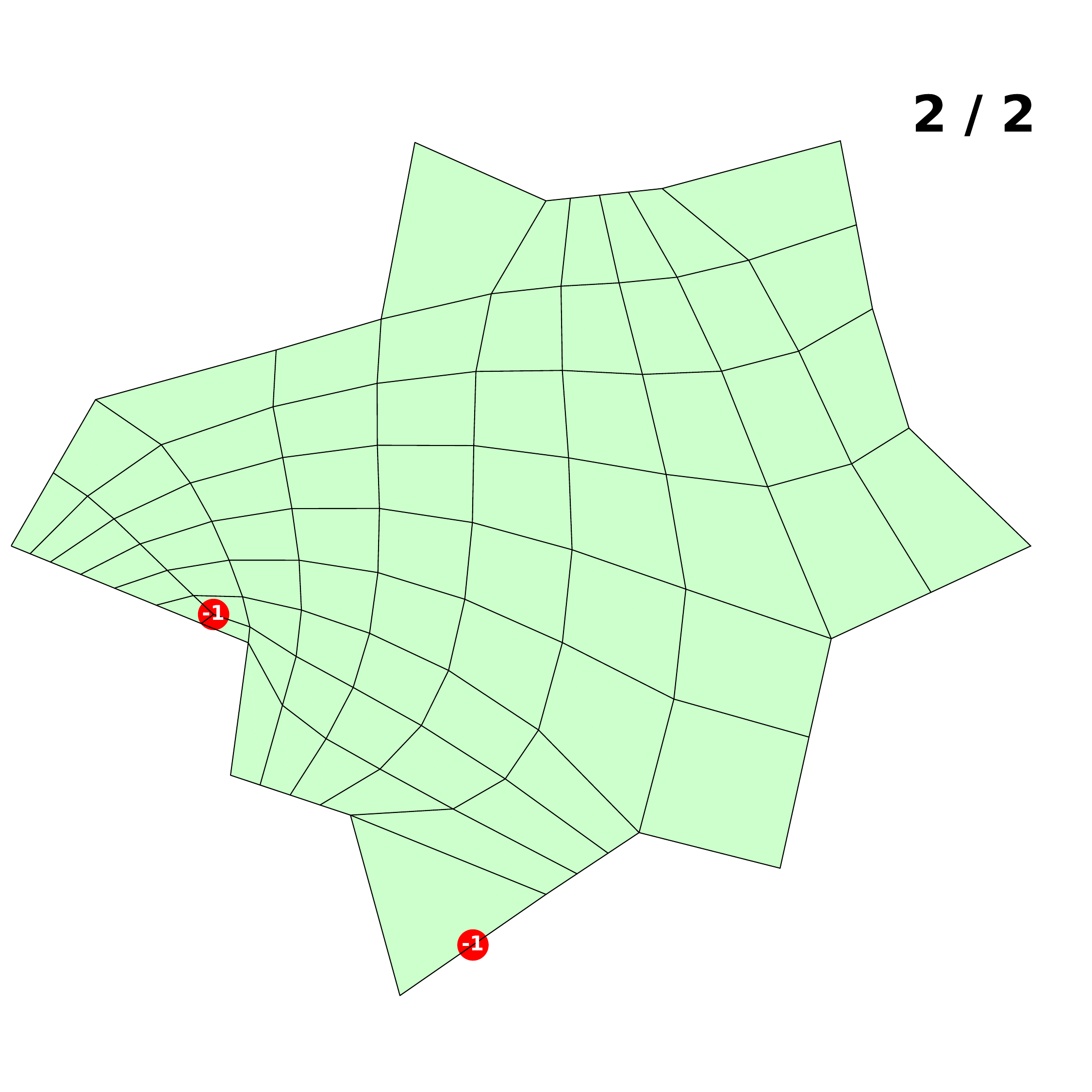}
    \caption{}
    \end{subfigure}
    \caption{The same agent as before is able to optimize on a 20-sided polygonal shape using 40 operations. (a) initial mesh, (b) intermediate mesh, (c) final mesh} \label{fig:quad-20-side}
\end{figure}

\Cref{fig:quad-10-side,fig:quad-20-side} show some example rollouts on various polygon sizes. By using our local template, the vector of probabilities representing action selection by the agent is always of the same size regardless of the size of the mesh, ensuring that the probability distribution is unaffected by mesh size. We can think of our template selection as a (fast) candidate retrieval process.

\Cref{fig:quad-showcase} shows zero-shot transfer to never before seen geometries like L-shape, star-shape, etc. The model is able to handle geometries with re-entrant corners and notches. In particular, we highlight the use of our approach in block decomposition of complex shapes into coarse quadrilateral elements. The global cleanup operation is particularly effective for this application. 

\begin{table}
\centering
\begin{tabular}{c|c|c}
     Polygon degree & Average & Standard deviation  \\ \hline
     10 - 20 & 0.98 & 0.03 \\
     20 - 30 & 0.97 & 0.06 \\
     30 - 40 & 0.94 & 0.14 \\
     40 - 50 & 0.91 & 0.13 \\
\end{tabular}
\caption{Evaluating the quadrilateral mesh agent on various sized random polygons. The agent was trained purely on 10 - 20 sided polygons, but is able to generalize to larger polygonal shapes with minor deterioration in performance. The agent was evaluated by picking the best of 10 rollouts per geometry, with the statistics computed over 100 randomly generated polygonal shapes. Using a fixed-sized local template enables stronger generalization to different sized meshes.} \label{tab:quad-generalize}
\end{table}

\begin{figure}
    \centering
    \begin{subfigure}[b]{.3\linewidth}
    \includegraphics[width=\linewidth]{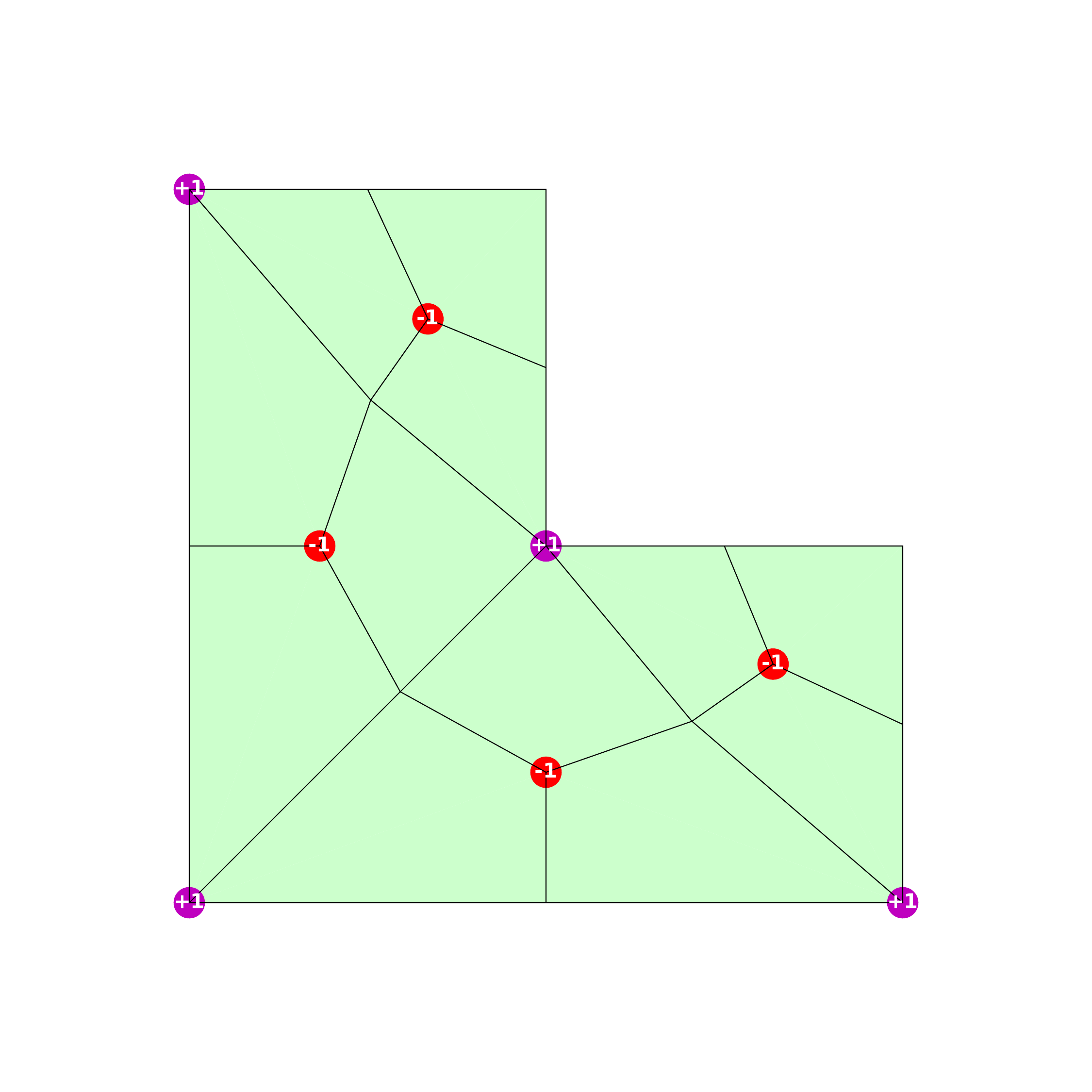}
    \end{subfigure}
    \begin{subfigure}[b]{.3\linewidth}
    \includegraphics[width=\linewidth]{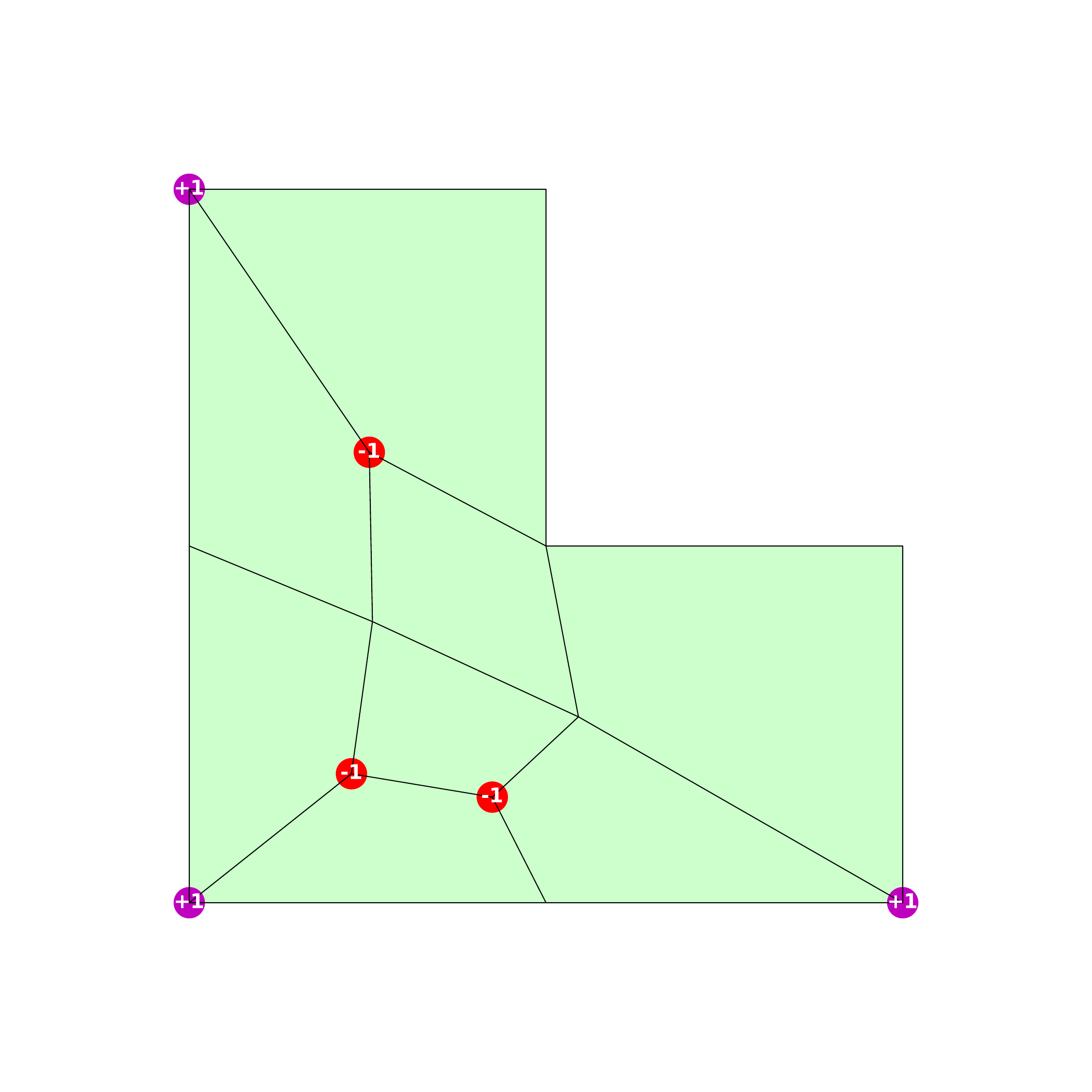}
    \end{subfigure}
    \begin{subfigure}[b]{.3\linewidth}
    \includegraphics[width=\linewidth]{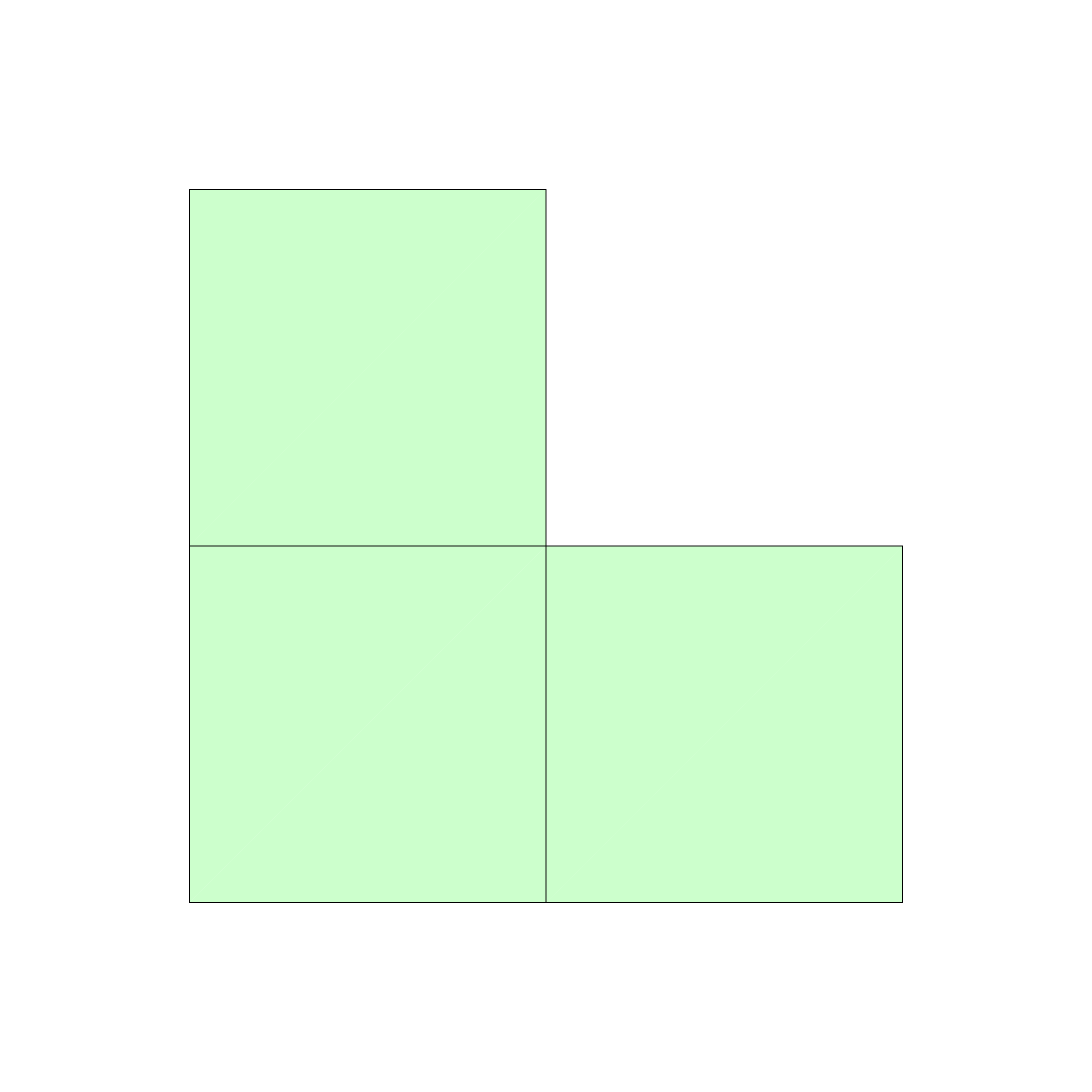}
    \end{subfigure}

    \begin{subfigure}[b]{.3\linewidth}
    \includegraphics[width=\linewidth]{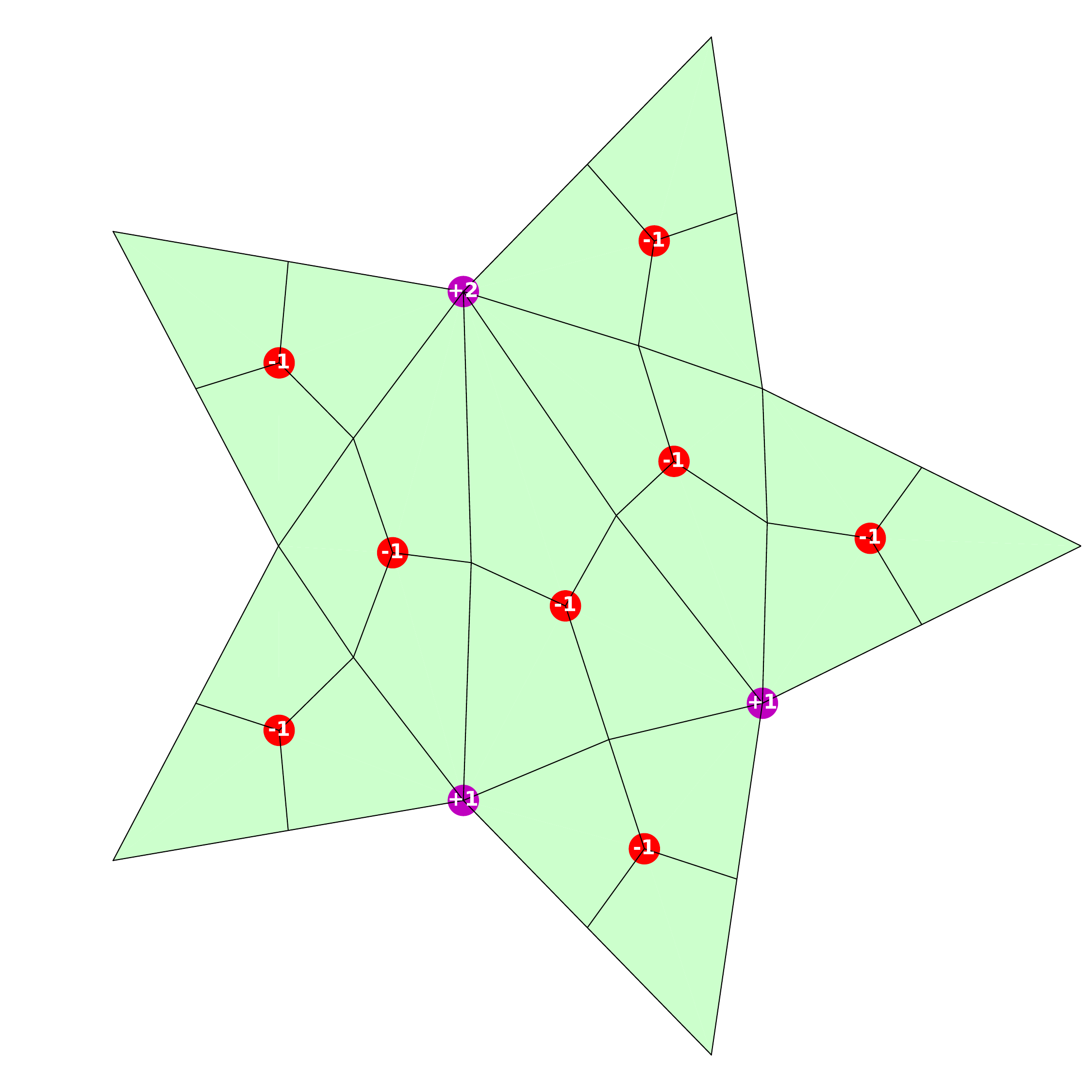}
    \end{subfigure}
    \begin{subfigure}[b]{.3\linewidth}
    \includegraphics[width=\linewidth]{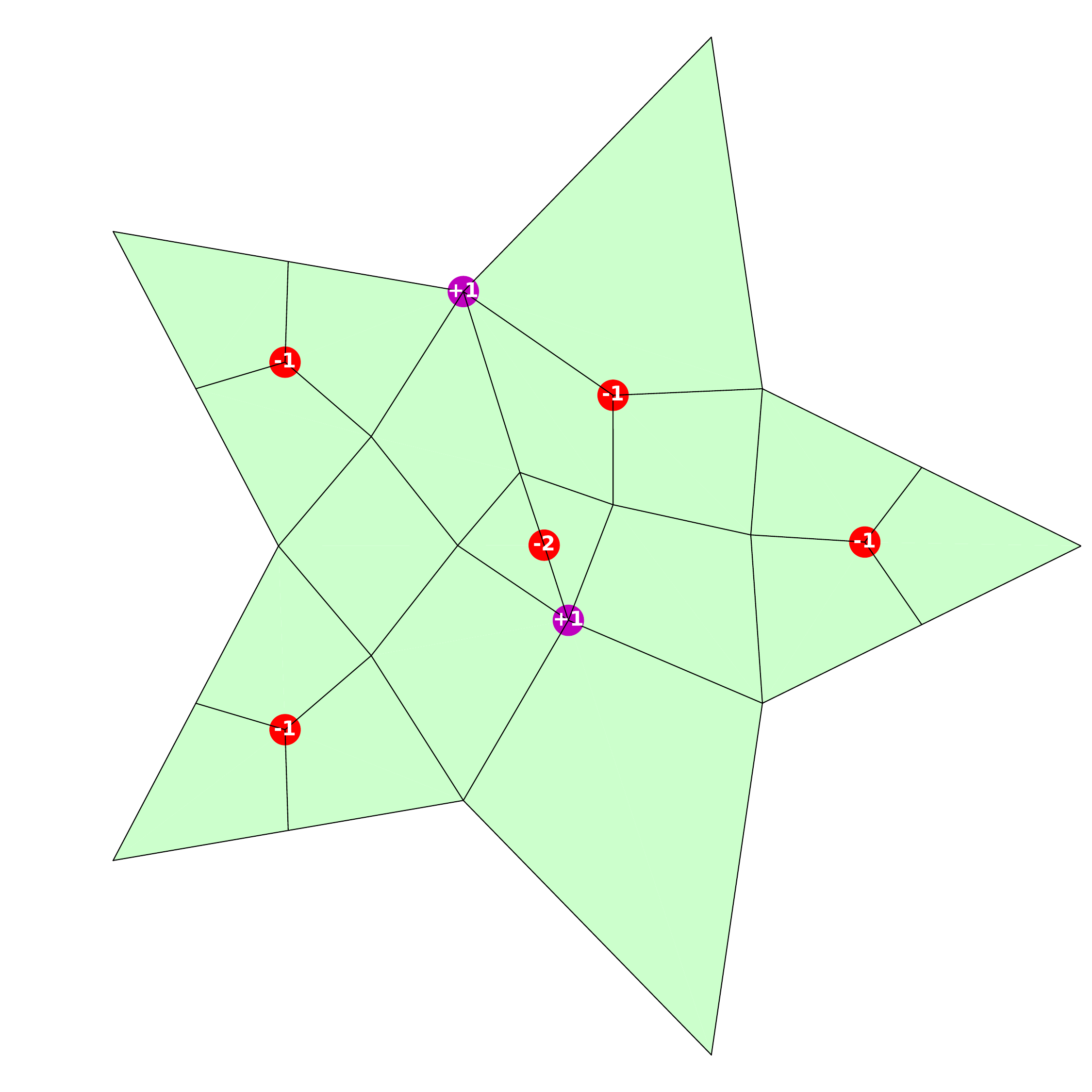}
    \end{subfigure}
    \begin{subfigure}[b]{.3\linewidth}
    \includegraphics[width=\linewidth]{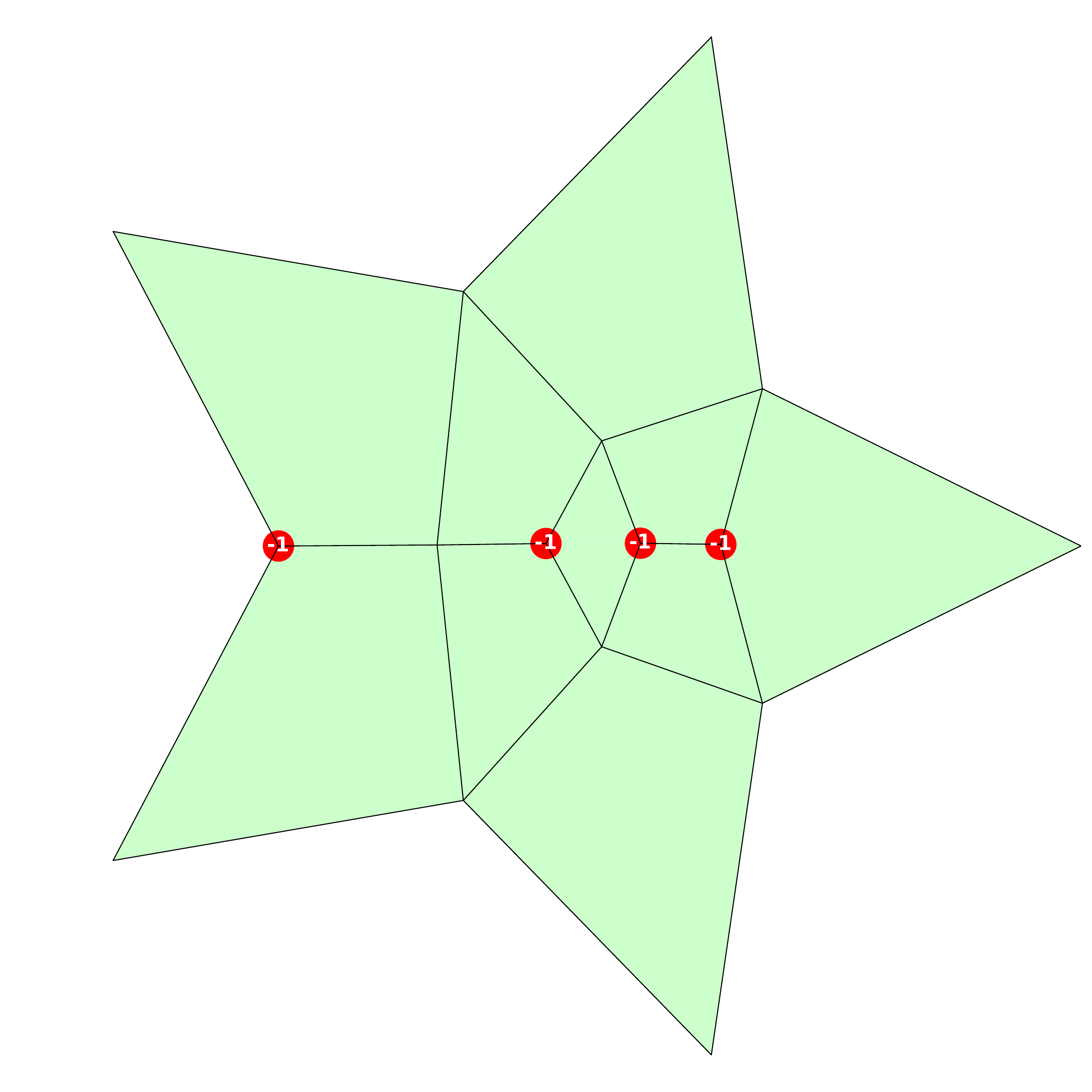}
    \end{subfigure}

    \begin{subfigure}[b]{.3\linewidth}
    \includegraphics[width=\linewidth]{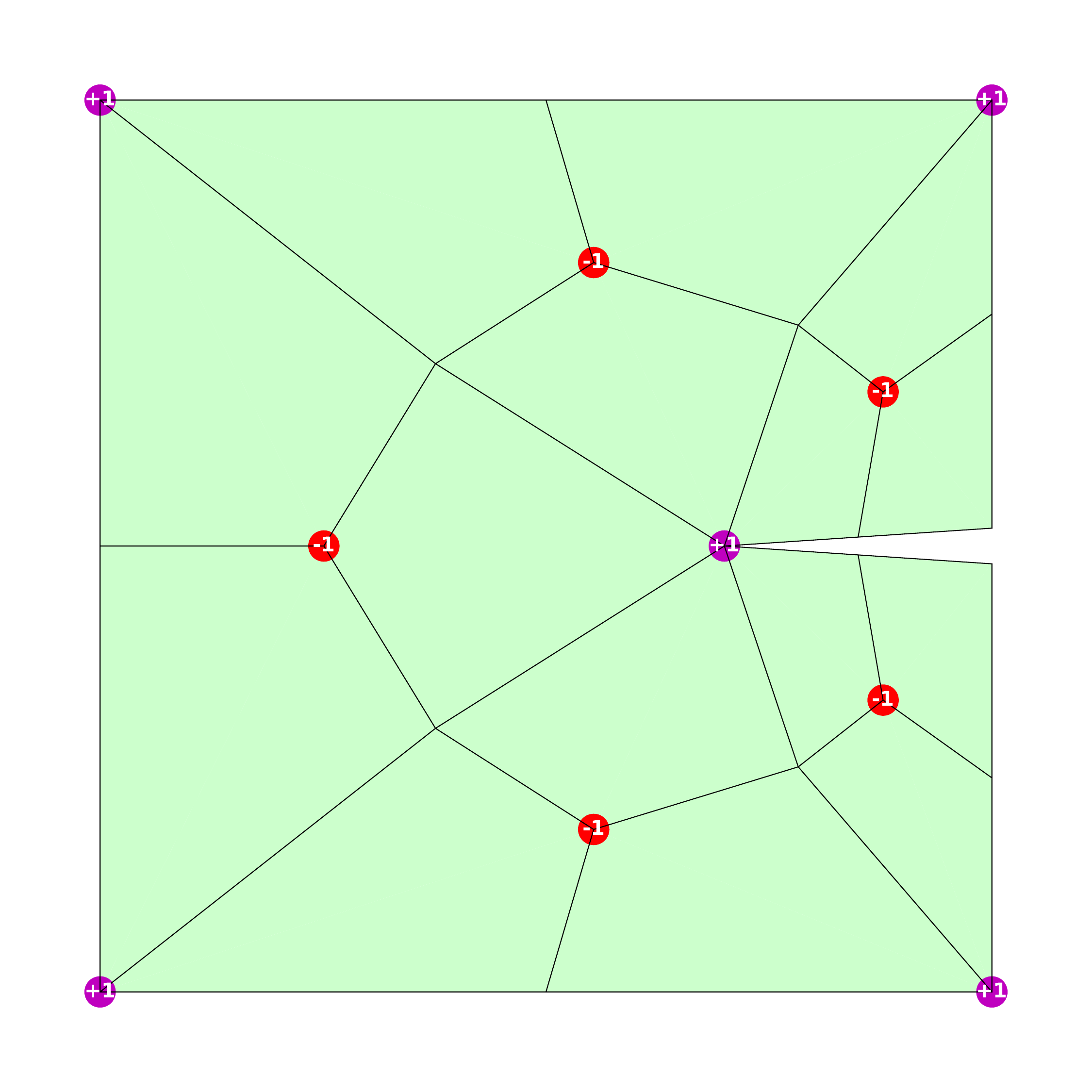}
    \end{subfigure}
    \begin{subfigure}[b]{.3\linewidth}
    \includegraphics[width=\linewidth]{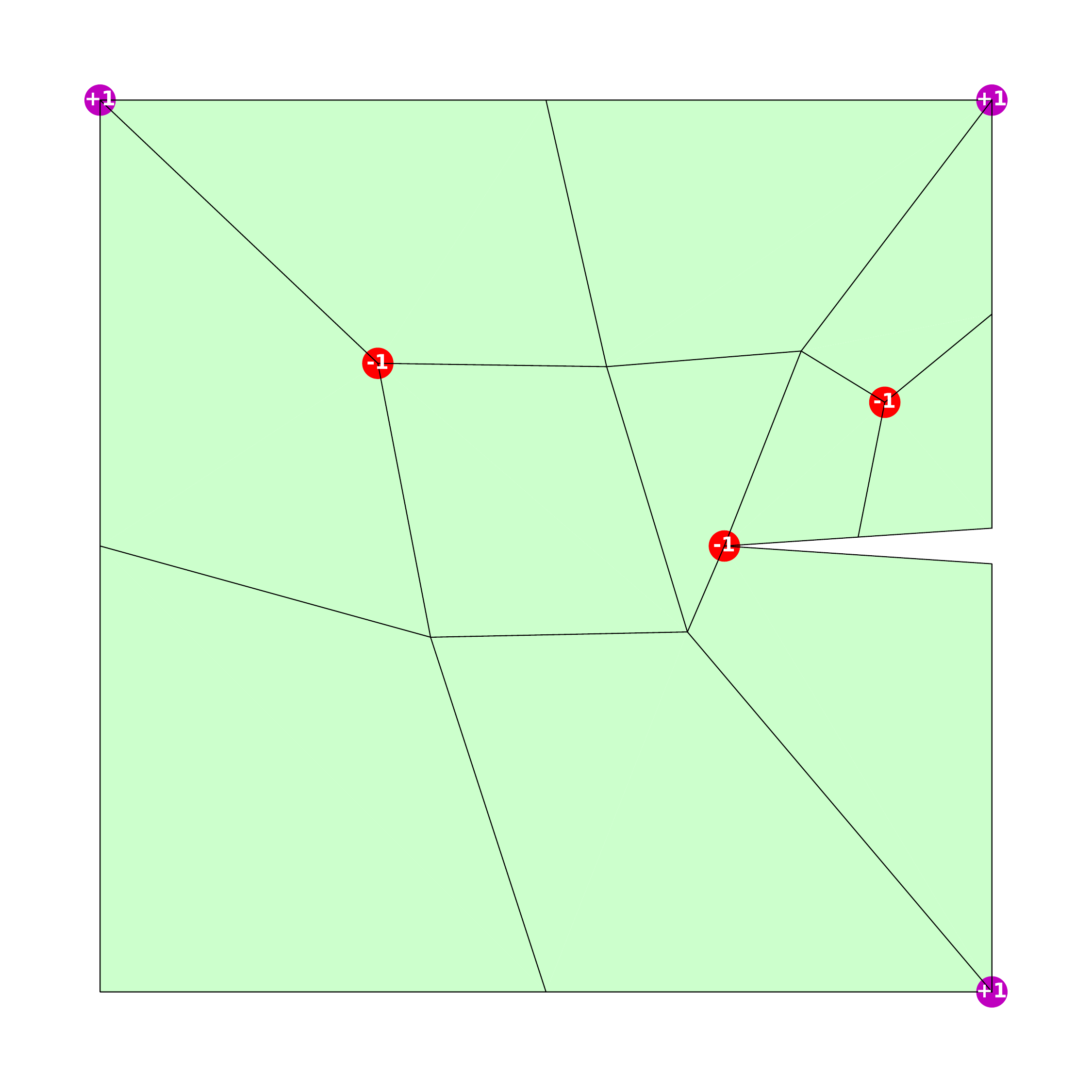}
    \end{subfigure}
    \begin{subfigure}[b]{.3\linewidth}
    \includegraphics[width=\linewidth]{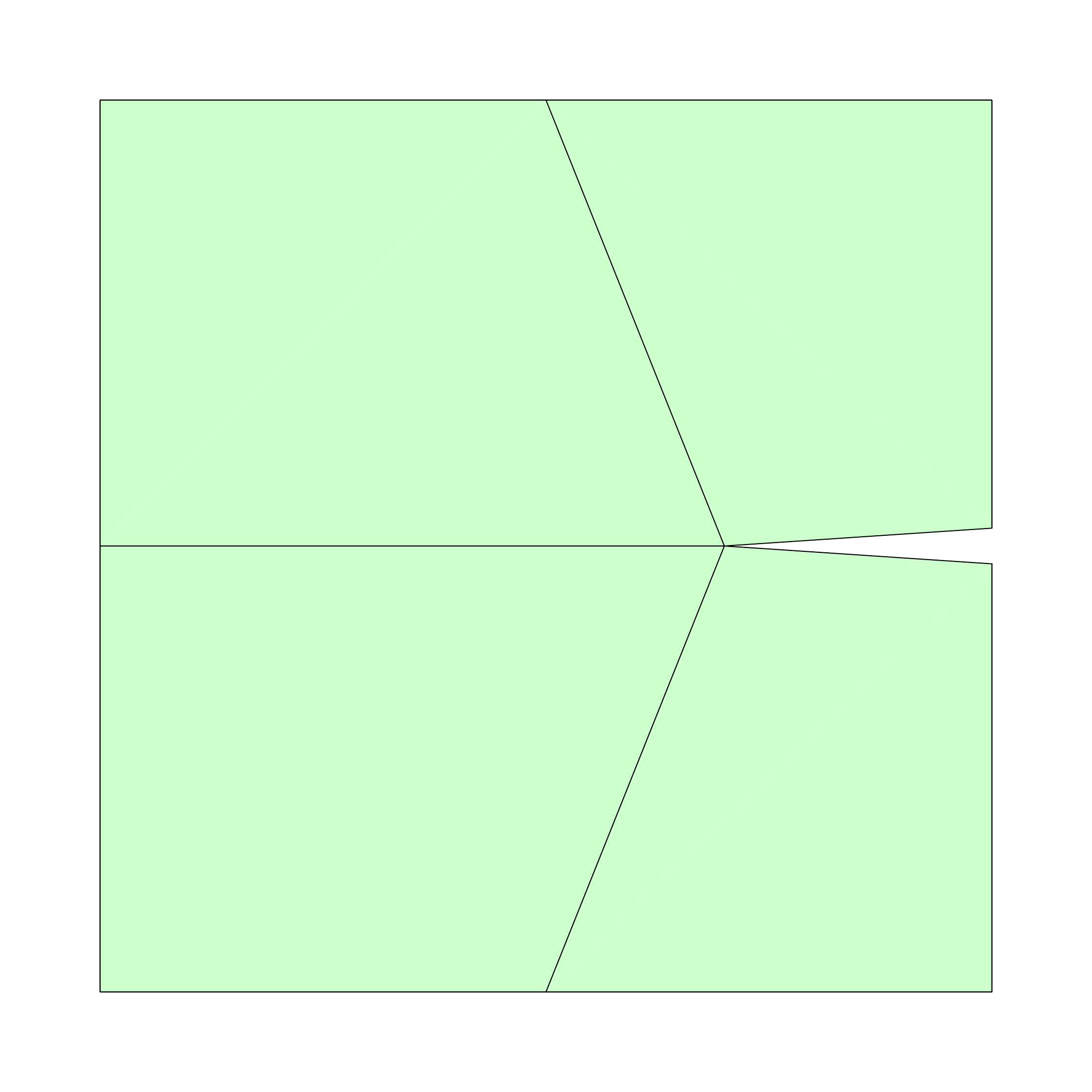}
    \end{subfigure}

    \begin{subfigure}[b]{.3\linewidth}
    \includegraphics[width=\linewidth]{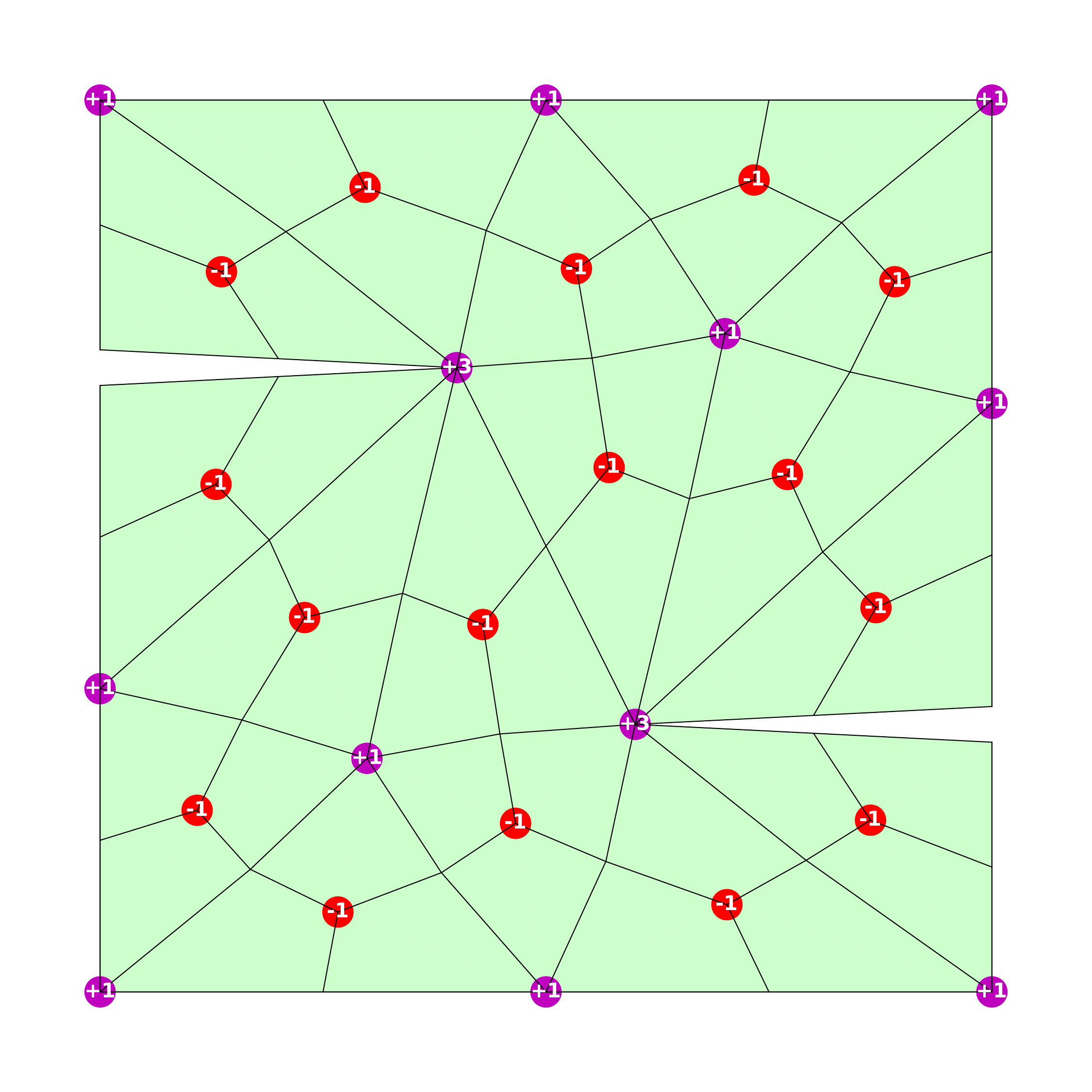}
    \end{subfigure}
    \begin{subfigure}[b]{.3\linewidth}
    \includegraphics[width=\linewidth]{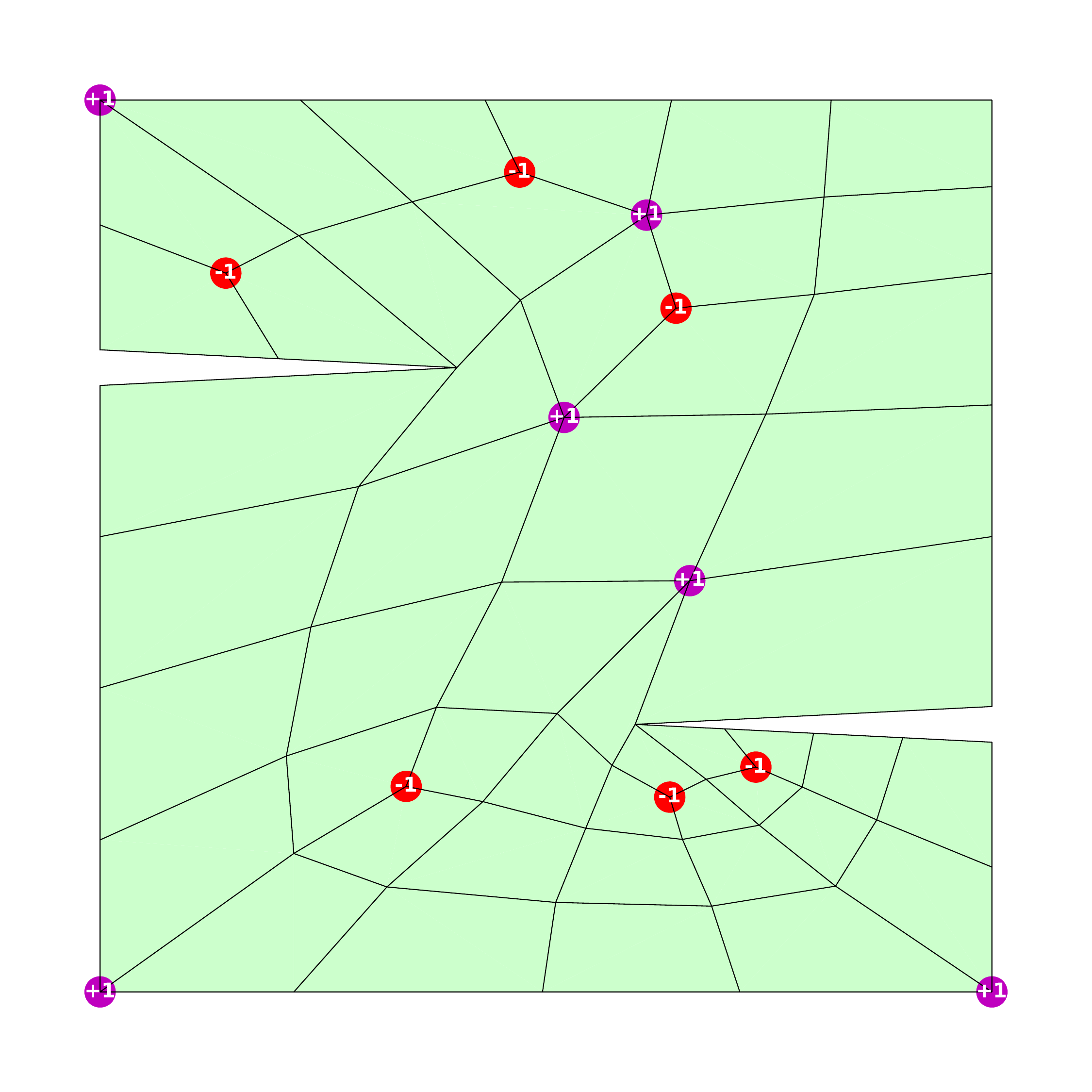}
    \end{subfigure}
    \begin{subfigure}[b]{.3\linewidth}
    \includegraphics[width=\linewidth]{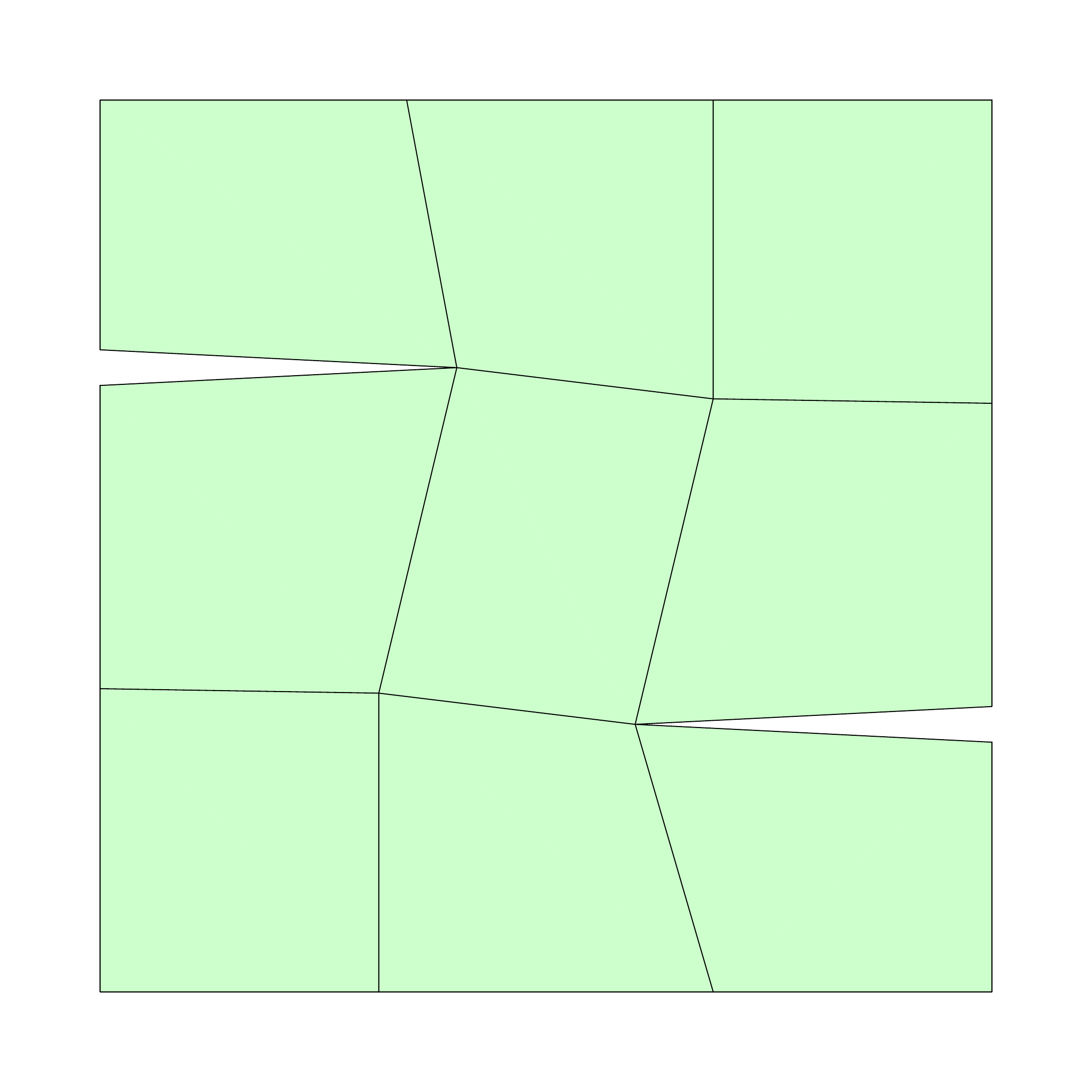}
    \end{subfigure}
\end{figure}
\begin{figure}[ht]\ContinuedFloat
    \begin{subfigure}[b]{.3\linewidth}
    \includegraphics[width=\linewidth]{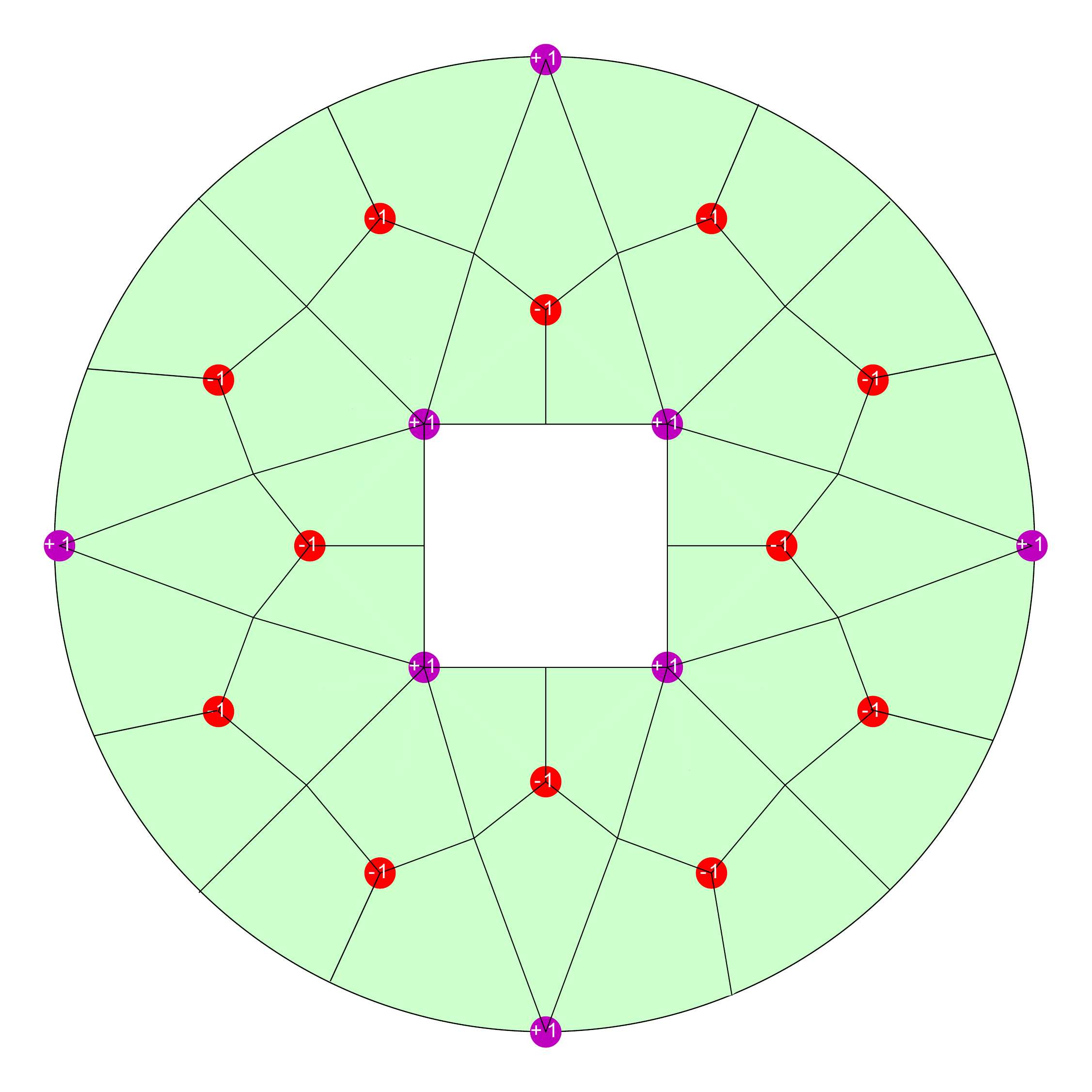}
    \end{subfigure}
    \begin{subfigure}[b]{.3\linewidth}
    \includegraphics[width=\linewidth]{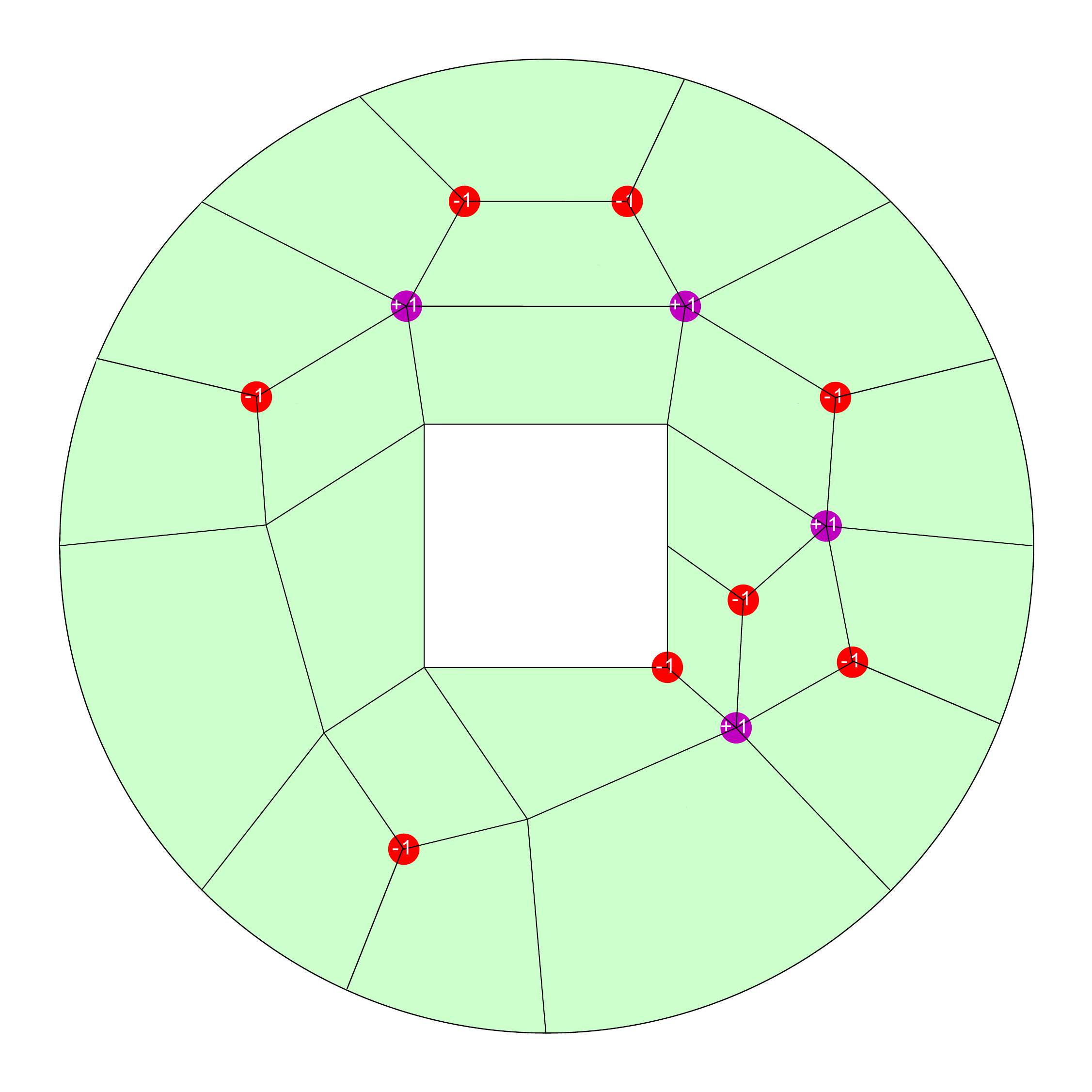}
    \end{subfigure}
    \begin{subfigure}[b]{.3\linewidth}
    \includegraphics[width=\linewidth]{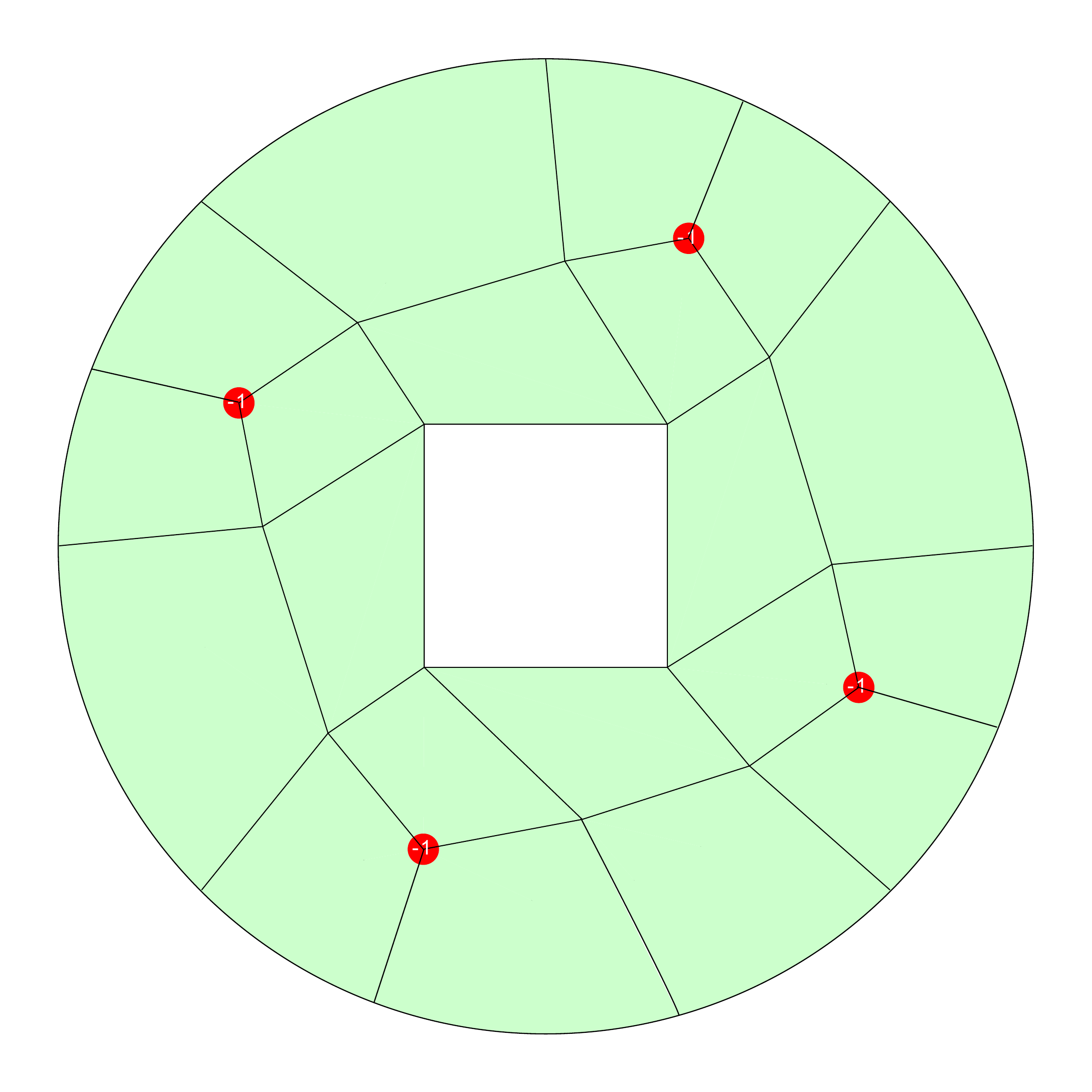}
    \end{subfigure}
    
    \caption{The quadrilateral mesh agent demonstrates zero-shot transfer on geometries that were not seen during training, including geometries with re-entrant corners like the star shape and the single and double-notch domains. Since our model is based purely on connectivity, we can directly transfer the model onto geometries with holes even though such geometries were never seen during training. We are particularly interested in coarse meshes representing block decompositions of more complex shapes. Notice that the star-shaped domain and the circular mesh with a square hole contain intrinsic irregularity that cannot be improved upon with our prescribed operations and heuristic. The cleanup operation is particularly useful in achieving coarse meshes. This is most evident in the single notch and double notch example in row 3 and 4.} \label{fig:quad-showcase}
\end{figure}

\section{Conclusions}

We presented here a method that learns to improve the connectivity of triangular and quadrilateral meshes through self-play reinforcement learning without any human input. A key contribution of this work is a parameterized method to generate a representation of the local topology in mesh neighborhoods. This enables appropriate selection of standard topological mesh editing operations which result in the reduction of irregular vertices in the mesh. Our method is built on the DCEL data-structure which allows the same framework to work on any planar 2D mesh with the discussion in this paper restricted to triangular and quadrilateral meshes.

There are several exciting directions of future research:

\begin{itemize}
    \item \textbf{Policy improvement with tree search:} our learned policy may be combined with e.g. Monte Carlo Tree Search (MCTS) \cite{coulom2006efficient} to efficiently search for optimal meshes for a specific geometry. The performance improvements that we observe from our naive best-of-k method in \cref{sec:tri-results,sec:quad-results} indicates that MCTS could be effective at improving the performance of our trained model. Such an approach would be similar to the AlphaZero \cite{silver2018general} system.
    \item \textbf{Optimizing for element quality:} to achieve this, our model would need to additionally receive geometric information (e.g. vertex coordinates) as input. This can be easily achieved by including the coordinates of vertices as part of the input features to our model (see \cref{sec:dcel-convolution}). We expect that the coordinates need to be normalized e.g. affine transform half-edges (and all vertices in its template) to a normalized coordinate system (e.g. $[0,1]$.)
    \item \textbf{Extension to 3D:} We expect that our method can leverage the equivalent of the half-edge data-structure in 3D \cite{dobkin1987primitives} to learn topological mesh editing operations on tetrahedral and hexahedral meshes. Determining optimal sequences of operations in 3D is highly challenging, and a self-learning method would have significant use.
\end{itemize}

A major advantage of artificial intelligence is its ability to discover heuristics that are too laborious and cumbersome for humans to identify, formulate, and prescribe. There are several areas in mesh generation where the automatic discovery of such heuristics can significantly aid engineers in their work. We hope that this paper demonstrates one such use-case.

\section*{Acknowledgments}

This work was supported in part by the Director, Office of Science, Office of Advanced Scientific Computing Research, U.S. Department of Energy under Contract No. DE-AC02-05CH11231.

\bibliographystyle{plain}
\bibliography{references}

\begin{thebibliography}{10}

\bibitem{akram2022structure}
Muhammad~Naeem Akram, Kaoji Xu, and Guoning Chen.
\newblock Structure simplification of planar quadrilateral meshes.
\newblock {\em Computers \& Graphics}, 109:1--14, 2022.

\bibitem{blacker1991paving}
Ted~D Blacker and Michael~B Stephenson.
\newblock Paving: A new approach to automated quadrilateral mesh generation.
\newblock {\em International journal for numerical methods in engineering},
  32(4):811--847, 1991.

\bibitem{bommes2013quad}
David Bommes, Bruno L{\'e}vy, Nico Pietroni, Enrico Puppo, Claudio Silva, Marco
  Tarini, and Denis Zorin.
\newblock Quad-mesh generation and processing: A survey.
\newblock In {\em Computer graphics forum}, volume~32, pages 51--76. Wiley
  Online Library, 2013.

\bibitem{coulom2006efficient}
R{\'e}mi Coulom.
\newblock Efficient selectivity and backup operators in monte-carlo tree
  search.
\newblock In {\em International conference on computers and games}, pages
  72--83. Springer, 2006.

\bibitem{daniels2008quadrilateral}
Joel Daniels, Cl{\'a}udio~T Silva, Jason Shepherd, and Elaine Cohen.
\newblock Quadrilateral mesh simplification.
\newblock {\em ACM transactions on graphics (TOG)}, 27(5):1--9, 2008.

\bibitem{diprete2023reinforcement}
Benjamin~C DiPrete, Rao~V Garimella, Cristina~Garcia Cardona, and Navamita Ray.
\newblock Reinforcement learning for block decomposition of cad models.
\newblock {\em arXiv preprint arXiv:2302.11066}, 2023.

\bibitem{dobkin1987primitives}
David~P Dobkin and Michael~J Laszlo.
\newblock Primitives for the manipulation of three-dimensional subdivisions.
\newblock In {\em Proceedings of the third annual symposium on Computational
  geometry}, pages 86--99, 1987.

\bibitem{docampo2019towards}
Julia Docampo-Sanchez and Robert Haimes.
\newblock Towards fully regular quad mesh generation.
\newblock In {\em AIAA Scitech 2019 Forum}, page 1988, 2019.

\bibitem{docampo2020regularization}
Julia Docampo-S{\'a}nchez and Robert Haimes.
\newblock A regularization approach for automatic quad mesh generation.
\newblock {\em 28th International Meshing Roundtable. Zenodo}, 2020.

\bibitem{dyedov2015ahf}
Vladimir Dyedov, Navamita Ray, Daniel Einstein, Xiangmin Jiao, and Timothy~J
  Tautges.
\newblock Ahf: Array-based half-facet data structure for mixed-dimensional and
  non-manifold meshes.
\newblock {\em Engineering with Computers}, 31:389--404, 2015.

\bibitem{klingner2007aggressive}
Bryan~Matthew Klingner and Jonathan~Richard Shewchuk.
\newblock Aggressive tetrahedral mesh improvement.
\newblock In {\em Proceedings of the 16th international meshing roundtable},
  pages 3--23. Springer, 2007.

\bibitem{ledoux2010topological}
Franck Ledoux and Jason Shepherd.
\newblock Topological modifications of hexahedral meshes via sheet operations:
  a theoretical study.
\newblock {\em Engineering with Computers}, 26:433--447, 2010.

\bibitem{mark2008computational}
de~Berg Mark, Cheong Otfried, van~Kreveld Marc, and Overmars Mark.
\newblock {\em Computational geometry algorithms and applications}.
\newblock Spinger, 2008.

\bibitem{owen1999qmorph}
Steven~J Owen, Matthew~L Staten, Scott~A Canann, and Sunil Saigal.
\newblock Q-morph: an indirect approach to advancing front quad meshing.
\newblock {\em International journal for numerical methods in engineering},
  44(9):1317--1340, 1999.

\bibitem{pan2023reinforcement}
Jie Pan, Jingwei Huang, Gengdong Cheng, and Yong Zeng.
\newblock Reinforcement learning for automatic quadrilateral mesh generation: A
  soft actor--critic approach.
\newblock {\em Neural Networks}, 157:288--304, 2023.

\bibitem{pan2021self}
Jie Pan, Jingwei Huang, Yunli Wang, Gengdong Cheng, and Yong Zeng.
\newblock A self-learning finite element extraction system based on
  reinforcement learning.
\newblock {\em AI EDAM}, 35(2):180--208, 2021.

\bibitem{peraire1987adaptive}
Jaime Peraire, Morgan Vahdati, Ken Morgan, and Olgierd~C Zienkiewicz.
\newblock Adaptive remeshing for compressible flow computations.
\newblock {\em Journal of computational physics}, 72(2):449--466, 1987.

\bibitem{remacle2012blossom}
J.-F. Remacle, J.~Lambrechts, B.~Seny, E.~Marchandise, A.~Johnen, and
  C.~Geuzainet.
\newblock Blossom-{Q}uad: a non-uniform quadrilateral mesh generator using a
  minimum-cost perfect-matching algorithm.
\newblock {\em Internat. J. Numer. Methods Engrg.}, 89(9):1102--1119, 2012.

\bibitem{schulman2017proximal}
John Schulman, Filip Wolski, Prafulla Dhariwal, Alec Radford, and Oleg Klimov.
\newblock Proximal policy optimization algorithms.
\newblock {\em arXiv preprint arXiv:1707.06347}, 2017.

\bibitem{shewchuk2002delaunay}
Jonathan~Richard Shewchuk.
\newblock Delaunay refinement algorithms for triangular mesh generation.
\newblock {\em Computational geometry}, 22(1-3):21--74, 2002.

\bibitem{shewchuk2002two}
Jonathan~Richard Shewchuk.
\newblock Two discrete optimization algorithms for the topological improvement
  of tetrahedral meshes.
\newblock {\em Unpublished manuscript}, 65:2--7, 2002.

\bibitem{silver2018general}
David Silver, Thomas Hubert, Julian Schrittwieser, Ioannis Antonoglou, Matthew
  Lai, Arthur Guez, Marc Lanctot, Laurent Sifre, Dharshan Kumaran, Thore
  Graepel, et~al.
\newblock A general reinforcement learning algorithm that masters chess, shogi,
  and go through self-play.
\newblock {\em Science}, 362(6419):1140--1144, 2018.

\bibitem{sluzalec2023quasi}
Tomasz S{\l}u{\.z}alec, Rafa{\l} Grzeszczuk, Sergio Rojas, Witold Dzwinel, and
  Maciej Paszy{\'n}ski.
\newblock Quasi-optimal hp-finite element refinements towards singularities via
  deep neural network prediction.
\newblock {\em Computers \& Mathematics with Applications}, 142:157--174, 2023.

\bibitem{sutton2018reinforcement}
Richard~S Sutton and Andrew~G Barto.
\newblock {\em Reinforcement learning: An introduction}.
\newblock MIT press, 2018.

\bibitem{tarini2010practical}
Marco Tarini, Nico Pietroni, Paolo Cignoni, Daniele Panozzo, and Enrico Puppo.
\newblock Practical quad mesh simplification.
\newblock In {\em Computer Graphics Forum}, volume~29, pages 407--418. Wiley
  Online Library, 2010.

\bibitem{tautgesa2003topology}
Timothy~J Tautgesa and Sarah~E Knoopb.
\newblock Topology modification of hexahedral meshes using atomic dual-based
  operations.
\newblock {\em algorithms}, 11:12, 2003.

\bibitem{vinyals2015pointer}
Oriol Vinyals, Meire Fortunato, and Navdeep Jaitly.
\newblock Pointer networks.
\newblock {\em Advances in neural information processing systems}, 28, 2015.

\bibitem{yang2023reinforcement}
Jiachen Yang, Tarik Dzanic, Brenden Petersen, Jun Kudo, Ketan Mittal, Vladimir
  Tomov, Jean-Sylvain Camier, Tuo Zhao, Hongyuan Zha, Tzanio Kolev, et~al.
\newblock Reinforcement learning for adaptive mesh refinement.
\newblock In {\em International Conference on Artificial Intelligence and
  Statistics}, pages 5997--6014. PMLR, 2023.

\end{thebibliography}

\end{document}